\documentclass[fleqn,usenatbib]{mnras}

\usepackage{siunitx}

\usepackage{graphicx}	
\usepackage{amsmath}	
\usepackage{amssymb}	
\usepackage{multirow}
\usepackage{xcolor}
\defcitealias{Yang2012ApJ}{Y12}
\usepackage{newtxtext,newtxmath}

\title[Emission from leptons in galaxy bubbles]{Multi-wavelength emission from leptonic processes in ageing galaxy bubbles}

\author[Owen \& Yang]{
Ellis R. Owen$^{1,2}$\thanks{E-mail: erowen@gapp.nthu.edu.tw (ERO), hyang@phys.nthu.edu.tw (HYKY)}, 
H.-Y. Karen Yang$^{1,2,3}$
\\
$^{1}$Institute of Astronomy, Department of Physics, National Tsing Hua University, Hsinchu, Taiwan (ROC)\\
$^{2}$Center for Informatics and Computation in Astronomy, National Tsing Hua University, Hsinchu, Taiwan (ROC)\\
$^{3}$Physics Division, National Center for Theoretical Sciences, Taipei, 106017, Taiwan (ROC)
}

\date{Accepted XXX. Received YYY; in original form ZZZ}

\pubyear{2020}

\begin{document}
\label{firstpage}
\pagerange{\pageref{firstpage}--\pageref{lastpage}}
\maketitle

\begin{abstract} 
The evolutionary behavior and multi-wavelength emission properties of bubbles around galaxies, such as the \textit{Fermi} bubbles of the Milky Way, is unsettled.  We perform 3D magnetohydrodynamical simulations to investigate the evolution of leptonic galaxy bubbles driven by a 0.3 Myr intense explosive outburst from the nucleus of Milky Way-like galaxies. Adopting an ageing model for their leptonic cosmic rays, we post-process our simulations to compute the multi-wavelength emission properties of these bubbles. We calculate the resulting spectra emitted from the bubbles
from radio frequencies to $\gamma$-rays, and 
construct emission maps in four energy bands to show the development of the spatial emission structure of the bubbles.
The simulated bubbles show a progression in their spectral properties as they age. In particular, the TeV $\gamma$-ray emission is initially strong and dominated by inverse Compton scattering, but falls rapidly after $\sim$ 1 Myr. By contrast, the radio synchrotron emission remains relatively stable and fades slowly over the lifetime of the bubble. 
Based on the emission properties of our post-processed simulations, we  demonstrate that $\gamma$-ray observations will be limited in their ability to detect galaxy bubbles, with only young bubbles around nearby galaxies being within reach. However, radio observations with e.g. the up-coming Square Kilometer Array, would be able to detect substantially older bubbles at much greater distances, and would be better placed to capture the evolutionary progression and diversity of galaxy bubble populations.  
\end{abstract}


%

\section{Introduction} 
\label{sec:intro} 

The \textit{Fermi} bubbles are a prominent feature in the $\gamma$-ray sky, and are arguably one of the most remarkable findings of the \textit{Fermi} $\gamma$-ray space telescope to date~\citep{Su2010ApJ, Dobler2010ApJ, Ackermann2014ApJ}. 
These giant $\gamma$-ray lobe structures extend up to \ang{50} symmetrically above and below the Galactic Centre (GC), with a total emission power of
  $>10^{37}{\rm erg}~{\rm s}^{-1}$ between $1 - 100~{\rm GeV}$. Their spectrum is hard (${\rm d}N_{\gamma}/{\rm d}E_{\gamma} \sim E_{\gamma}^{-2}$) with an exponential cut-off, and spatially-uniform, while their $\gamma$-ray surface brightness distribution is relatively flat, showing only a sharp boundary with marginal internal substructure~\citep{Yang2014AA, Ackermann2014ApJ, Narayanan2017MNRAS, Keshet2017ApJ}. 
  Despite gaining substantial attention since their discovery over a decade ago~\citep{Su2010ApJ}, 
  the origin and physical nature of the \textit{Fermi} bubbles remains unsettled, and several proposals have been discussed~\citep[see a review by][]{Yang2018Galax}. 
  \cite{Yang2012ApJ} (hereafter \citetalias{Yang2012ApJ}) considered their origins to lie in an intense explosive outburst of the central black hole at the Galactic Centre of the Milky Way, {Sgr A*} a few Myr ago (perhaps indicative of previous active galactic {nucleus}, AGN, activity), with the present $\gamma$-ray emission arising predominantly through  inverse Compton scattering of an energetic non-thermal electron population in the remnant cocoons (the so-called `leptonic' model -- see also~\citealt{Su2010ApJ, Zubovas2011MNRAS, Su2012ApJ, Fujita2013ApJ}).
  Alternative scenarios have also been discussed, where the bubbles may arise from the confluence of a number of processes arising more gradually within the inner part of the Milky Way~\citep{Thoudam2013ApJ}. These could include tidal disruption events (TDEs) occurring at regular intervals of 10s to 100s kyr~\citep{Cheng2011ApJ}, or from the action of a bipolar galactic outflow driven by the ongoing intense Galactic Centre star-formation activity and/or the processes associated with Sgr A*~\citep{Lacki2014MNRAS}, with the resulting $\gamma$-ray glow instead arising from a resulting hadronic cosmic ray (CR) population interacting with an advected supply of entrained gas and clumps in the wind (the `hadronic' models -- see ~\citealt{Crocker2011PhRvL, Crocker2014ApJ, Crocker2015ApJ, Mou2014ApJ, Mou2015ApJ, Cheng2014ApJ, Cheng2015ApJ, Razzaque2018Galax}).
   
  Recently, purely hadronic models {extending to PeV energies}
  have been disfavoured by new data -- in particular, by the non-detection of the \textit{Fermi} bubbles in $\gamma$-rays between 1-100 TeV~\citep{Abeysekara2017ApJ, Fang2017PhRvD, Sherf2017ApJ}. However, purely leptonic or hybrid lepto-hadronic scenarios~\citep[e.g.][]{Crocker2015ApJ, AlvarezHurtado2019ICRC} both remain plausible. Indeed, radio~\citep{Heywood2019Natur} and X-ray emission~\citep{Snowden1997ApJ, Su2010ApJ, Ponti2019Natur, Predehl2020Natur}, and a long-known microwave haze~\citep{Finkbeiner2004ApJ, Planck2013AA, Sasaki2015ApJ} have all been found to correlate well spatially with the bubbles, pointing towards a substantial leptonic component. In such a scenario, a fast flow is necessary to sustain the uniformly hard spectrum throughout the extent of the bubbles~\citep[e.g.][]{Hooper2013PDU}, as cooling would be more rapid than in a hadron-dominated system and would lead to spectral steepening~\citep{Guo2012ApJ}. In the leptonic AGN jet prescription, for example, the bubble formation timescale must be relatively short (as low as 1.2 Myr, according to~\citetalias{Yang2012ApJ}). 
   {Such rapid bubble growth could drive fluid instabilities and form ripples in the bubble surfaces. These are, however, not observed -- though viscosity may be invoked to suppress them~\citet{Guo2012ApJ_b}.\footnote{Alternatively, ~\citetalias{Yang2012ApJ} considered that, in fact, visibly large-scale hydrodynamical instabilities would have insufficient time to develop, even under a rapid growth scenario, if the initial jet velocity is low.} Although it remains unclear how the high outflow velocities of the hot gas would influence the kinematics of the cold entrained clouds as probed by UV absorption lines~\citep{Bordoloi2017ApJ},} 
   overall, the leptonic model is largely supported by observations, including the features seen in the $\gamma$-ray, X-ray and microwave bands~\citep{Yang2012ApJ, Yang2013MNRAS, Yang2017ApJ}. The microwave haze, $\sim$100 GeV spectral cut-off and $\gamma$-ray spectral uniformity can also be accounted for~\citep{Yang2017ApJ, Yang2018Galax}. 
  
  In addition to the \textit{Fermi} bubbles of our own Galaxy, 
high-energy emission structures have also been identified in our neighbouring galaxy, M31 
  \cite[][]{Pshirkov2016MNRAS}. Although the M31 bubbles are comparable in power ($\sim 10^{37}~{\rm erg}~{\rm s}^{-1}$) and show evidence of a morphology similar to the \textit{Fermi} bubbles, they are slightly smaller in extent. More recently, it has also been noted that they appear to be embedded in a larger $\gamma$-ray emitting CR halo extending 100-200 kpc from the M31 centre~\citep{Recchia2021arXiv}.
  $\gamma$-ray structures, of $\sim$100s kpc in scale,  
  have also been detected from the radio lobes of Centaurus A~\citep[see][]{Abdo2010Sci}, and these could 
  arguably be another member of the family of \textit{Fermi}-bubble like phenomena around galaxies, but at a different stage of evolution and/or associated with a more violent astrophysical host. 
  Further to this, emission structures in other wavebands have also been found in
   a number of other galaxies in the form of X-ray emission lobes and halos. 
For example, the galaxy NGC 3079 was recently found to host X-ray emitting nuclear super-bubbles analogous to the \textit{Fermi} Bubbles \citep{Li2019ApJ}, 
   and a large-scale diffuse non-thermal hard X-ray halo was
   also identified around NGC~891 
 \citep{HodgesKluck2018ApJ}.  
These detections open up pressing new questions about how widespread such galaxy bubbles actually are, and whether these examples 
are unusual in some respect, or if these phenomena are ubiquitous throughout the galaxies of the Universe and develop into structures we are more familiar with in other wavebands as they age.

In this paper, we present a model for the evolution and expected multi-wavelength emission properties of \textit{Fermi} bubble structures as they evolve. We adopt the {bubble} model and methodology of~\citetalias{Yang2012ApJ}, and consider a leptonic scenario when calculating their emission properties. 
{This is the first study that explores the multi-wavelength non-thermal emission of galaxy bubbles and their long term evolution, and we focus on their emission signatures. Variation of model parameters and configurations will be investigated in future works (e.g. AGN activity, host galaxy properties, bubble environment, and the particle composition and energy spectrum of the CRs in the bubble). 
Modeling expected spectral signatures for bubbles in different evolutionary stages can provide information about the most valuable spectral signatures and, hence, the most appropriate energy bands to probe the physical conditions within bubbles around distant galaxies.
This can increase the useful reach of future observational studies, offering the potential to gather broader data to inform detailed population analyses. Such analyses will be crucial to understand the nature and evolution of galaxy bubbles, and how those around our own Galaxy fit into a larger galaxy bubble demographic.}

We arrange this paper as follows: In section~\ref{sec:section2} we outline our numerical model and initial conditions. In section~\ref{sec:section3}, we introduce the relevant particle interactions and our multi-wavelength emission calculations under a leptonic scenario. In section~\ref{sec:section4}, we present our results and show the multi-wavelength emission of ageing high-energy bubbles from galaxies, and discuss observational prospects with current and near-future instruments in section~\ref{sec:section5}. Finally, we summarize our results and draw conclusions in section~\ref{sec:section6}.
 
\section{Galaxy Bubble Model}
\label{sec:section2}

\subsection{Numerical approach}
\label{sec:numerical_method}

We model the evolution of galaxy bubbles using 3D magneto-hydrodynamical (MHD) simulations, with an implementation broadly following that described in~\citetalias{Yang2012ApJ} (see also~\citealt{Yang2013MNRAS}), which we refer the reader to for the detailed code description, initial conditions and numerical techniques. 
Under this prescription, the bubble inflation is driven by an initial 0.3 Myr injection of CR energy from the centre of the model galaxy, corresponding to an intensive explosive outburst from a period of AGN activity, with total injected energy $E_{\rm j} = 3.13\times 10^{57} \;\! {\rm erg}$ per jet. The subsequent development of the bubble is simulated using FLASH4~\citep{Fryxell2000ApJS, Dubey2008PhST}, an adaptive mesh refinement (AMR) code. 
Compared to the~\citetalias{Yang2012ApJ} set-up, we make a number of adjustments. In particular, 
 given that the longer-term evolution of the emerging bubbles is of interest in this work,
we adopt a larger simulation box than used in~\citetalias{Yang2012ApJ}, with 75 kpc on each side. Moreover, we introduce a progressive grid refinement process, undertaken in parts of the grid where the CR energy density exceeds $10^{-10}~{\rm erg}\;\!{\rm cm}^{-3}$ in each simulation step, with resolutions ranging from 4.7 kpc (coarsest) to {0.6 kpc} (finest). We set diode boundary conditions to allow for the outflow of gas and CRs but to prevent inflows into the simulation domain, and solve the MHD equations using the directionally unsplit staggered mesh solver~\citep{Lee2009JCoPh, Lee2013JCoPh} which ensures divergence-free magnetic fields. 
The injection of CRs is performed according to the same model set-up and jet parameters as previously used, in~\citetalias{Yang2012ApJ}. These parameters were carefully chosen to match the observed morphology of the Milky Way's \textit{Fermi} bubbles at 1.2 Myr, the limb-brightened intensity distribution in the \textit{ROSAT} X-ray 1.5 keV map and the gas temperature inside the bubbles inferred from the X-ray line ratios~\citep{Miller2013ApJ}. 
{Such a Galactic-based approach is adopted as a reference in this first study,
which puts focus on the multi-wavelength non-thermal emission signatures of galaxy bubbles and their long term evolution. Other model configurations and parameter choices (e.g., AGN activity, galaxy properties) will be explored in future works.} 

\subsubsection{Cosmic rays}

Within the MHD simulation, our treatment of CRs also follows that of~\citetalias{Yang2012ApJ}. These are modeled as a {second} fluid, for which the CR pressure evolution is solved directly. The CRs are advected with the thermal gas which, in turn, is modeled to react to gradients in the CR pressure,\footnote{Similar approaches have been adopted in other works -- e.g. see~\citet{Yu2021MNRAS}, where the impact of CRs on galactic outflows is modeled numerically.} {and diffuse along magnetic field lines}. Distinction between leptonic and hadronic CRs is not made at this stage, and related CR cooling/heating processes, interactions and energy spectra, are neglected {within the MHD simulations}. Although cooling processes and the CR composition will be explored in future studies, we note that their impact is likely to be subsidiary to the bubble properties of interest here, i.e. their 
macroscopic evolutionary behavior and survival time. For example, for a typical power-law CR {energy} spectrum, the bulk of the energy density would lie between 1 and 10 GeV. Even if densities, magnetic fields and radiation fields were as strong in our simulated bubbles as a star-forming galaxy's interstellar medium, cooling/absorption timescales of CRs (whether they are predominantly electrons or protons) would exceed several 10s Myr at $z=0$~\citep[e.g.][]{Owen2019AA} -- i.e. longer than the expected evolutionary timescale of the bubbles.\footnote{Note that this approximation would not hold at higher redshifts, where inverse Compton cooling of electrons against cosmological microwave background radiation would substantially suppress the CR energy density within the bubble, if the CRs were predominantly comprised of electrons.} The effects of CR composition and exact energy spectrum would instead manifest in the emission spectra of bubble structures (see section~\ref{sec:section3}), rather than their dynamical evolution.

In the MHD formulation of~\citetalias{Yang2012ApJ}, the CR evolution equation is written as
\begin{equation}
\frac{\partial e_{\rm CR}}{\partial t} + \nabla \cdot (e_{\rm CR} {\boldsymbol v}) = - p_{\rm CR}\nabla \cdot {\boldsymbol v} + \nabla \cdot ({\boldsymbol \kappa} \cdot \nabla e_{\rm CR}) \ ,
\end{equation}
where $e_{\rm CR}$ is the CR energy density, ${\boldsymbol v}$ is the velocity of the thermal fluid, $p_{\rm CR}$ is the CR pressure, and ${\boldsymbol \kappa}$ is the CR diffusion tensor. The diffusion term may be expressed as
\begin{equation}
\nabla \cdot ({\boldsymbol \kappa} \cdot \nabla e_{\rm CR}) = \nabla \cdot (\kappa_{||} \hat{\boldsymbol b} \hat{\boldsymbol b} \cdot \nabla e_{\rm CR}) + \nabla \cdot ( \kappa_{\perp} ({\boldsymbol I} - \hat{\boldsymbol b} \hat{\boldsymbol b}) \cdot \nabla e_{\rm CR})
\end{equation}
in the presence of a magnetic field, where $\kappa_{||}$ is the diffusion coefficient parallel to the magnetic field, $\kappa_{\perp}$ is the perpendicular component, and $\hat{\boldsymbol b} = {\boldsymbol B}/\langle|{\boldsymbol B}|\rangle$ is a unit vector in the direction of the local magnetic field. In our model, we expect that $\kappa_{||}$ and $\kappa_{\perp}$ are defined with respect to a mean, as the magnetic field presumably has structure on scales below our numerical resolution. However, for the purposes of this work, we adopt the approximation that the true field is oriented along the mean field (as, typically, $\kappa_{\perp} \ll \kappa_{||}$ -- see, e.g.~\citealt[][]{Ensslin2003AA}) and so set $\kappa_{\perp} = 0$. \citetalias{Yang2012ApJ} found that adopting such an approximation does not significantly change the results of the simulation. For $\kappa_{||}$, we adopt a value of $4.0\times 10^{28}\;\!{\rm cm}^2\;\!{\rm s}^{-1}$. Typically the effective isotropic diffusion coefficient in the Galaxy is constrained by observations to be $(3-5)\times 10^{28}\;\!{\rm cm}^2\;\!{\rm s}^{-1}$~\citep{Strong2007ARNPS}. In a tangled magnetic field, this value could be suppressed compared to $\kappa_{||}$ along magnetic field lines~\citep{Tao1995MNRAS}, thus our choice of value within this range is intended to represent a conservative estimate for $\kappa_{||}$.

\subsubsection{Initial halo magnetic field}

We consider a randomly-oriented, tangled initial magnetic field for the host galaxy's halo, into which the simulated galaxy bubbles expand (again, this approach follows~\citetalias{Yang2012ApJ}, to which the reader is referred for further details). 
This initial choice is adopted to avoid imposing any particular shape or configuration on the host galaxy's magnetic field and is intended instead to demonstrate the impact that a range of plausible magnetic field structures may have on the bubbles' development.
We take this approach due to the great uncertainties associated with the configuration of the halo magnetic field around galaxies (including that in our own galactic neighbourhood -- see, e.g. ~\citealt{Brown2010ASPC, Haverkorn2015ASSL}). 
As with the previous work, we investigate simulations where two different magnetic field coherence lengths are initially adopted. In the first case (which we hereafter refer to as `Run A', and is our baseline simulation), we compute a tangled field with coherence length $\ell_c =1$ kpc, being representative of the regular magnetic field of the host galaxy. 
Although the magnetic field in the Milky Way's disk -- and, presumably, in the host galaxies we model here -- is comprised of the sum of several components~\citep{Beck2015AArev} including a large-scale regular field and small-scale turbulent field, only the regular field's structure would be resolvable in our simulations.\footnote{The turbulent component of an interstellar galactic magnetic field would typically have a coherence length of $\sim$ 5-50 pc~\citep[see][for reviews]{Noutsos2012SSRv, Beck2015AArev, Haverkorn2015ASSL}, and so would not be resolved in our simulations.} In the second case (hereafter `Run B'), we compute a magnetic field with a coherence length of $\ell_c = 9$ kpc. This is intended to be reflective of the possible scale of magnetic structure in a galactic halo, which are found to be larger than 1 kpc~\citep{Krause2020AA}. Our exact choice of $\ell_c = 9$ kpc is expected to be small enough to produce a magnetic field that is sufficiently non-uniform to the developing bubble throughout its later evolution, but large enough to yield qualitatively different and distinguishable results compared to the $\ell_c =1$ kpc coherence length magnetic field scenario. 
In both cases, we normalize the magnetic field to have an averaged strength of 1 $\mu$G~\citep{Brown2010ASPC}.
We run all simulations with both initial magnetic field models to cover the extent of the range of possible magnetic field structures that may be encountered by an expanding bubble throughout its evolution.\footnote{These magnetic field initial configurations, together with our choice of diffusion coefficient, reflect the simulation Runs D and I of~\citetalias{Yang2012ApJ}.} {We note that we find the exact choice of $\ell_c$ is not of great importance to the resulting emitted spectra from a bubble, however it does impact on the spatial structure of the small-scale emission patterns from the system (see section~\ref{sec:section4} for details).} 

\subsubsection{Initial galaxy and halo gas model}
\label{sec:gal_input_model}

The initial galaxy and halo gas distribution model follows~\citet{Guo2012ApJ}, where a hot isothermal gas halo is set to be in hydrostatic equilibrium with a fixed Galactic potential. The potential includes contributions from the halo, disk and bulge, following the approach of~\cite{Helmi2001MNRAS}, where model parameters are set to the same as those used in~\citetalias{Yang2012ApJ}. 
The host galaxy model is positioned at the center of the simulation domain, where the initial gas temperature and density are set to be $T_0 = 2\times 10^6\;\!{\rm K}$ and $\rho = 3.88 \times 10^{-24}\;\!{\rm g}\;\!{\rm cm}^{-3}$ (corresponding to a thermal electron density of $n_{\rm e, 0} = 2 \;\!{\rm cm}^{-3}$) to match the observed hot gas density profile of~\cite{Miller2013ApJ} (see also~\citealt{Miller2015ApJ}).

\subsubsection{Radiation field}
\label{sec:gal_rad_field}

In a leptonic scenario, the emission from galaxy bubbles is attributed to energetic electrons. 
{In our simulations, the total CR energy density is degenerate with the thermal energy density (see also \citetalias{Yang2012ApJ}). We introduce a parameter $f_{\rm emit}$ which serves as a convenient scaling for the required number of CR electrons to match the bubble emission to a particular level. For consistency with previous work, we set this to $f_{\rm emit}=0.003$ so that other parameters may be maintained at values used earlier, in ~\cite{Yang2017ApJ}. The parameter $f_{\rm emit}$ is therefore not physical in our model and, from the degeneracy between CR energy density and thermal energy density, an alternative set of model parameters could be found where $f_{\rm emit}=1$, but where the physical results are identical to those obtained with our current parameter choices. We consider that the physical CR energy density is given by $f_{\rm emit} e_{\rm CR}$. These CRs undergo cooling and, hence, drive the leptonic emission in our model. The remaining fraction $(1-f_{\rm emit}) e_{\rm CR}$ does not cool (but acts as a dominating dynamical component).}\footnote{{Larger or smaller choices of $f_{\rm emit}$ with other parameters unchanged would lead to a proportional change in the intensity of the leptonic emission from the bubble {and the exact choice would therefore impact the detection prospects for bubbles located around other galaxies (see section~\ref{sec:obs_implications})}. 
The spectrum and spatial distribution of the emission would not be affected.}}

The CR electrons would cool via synchrotron, adiabatic, bremsstrahlung and inverse Compton processes, with their cooling rates and (where applicable) corresponding emission spectra being governed by the thermal gas properties, the magnetic and radiation field in the bubble, and the bubble expansion rate. While the macroscopic growth of the bubble and its thermal gas properties and magnetic field are computed self-consistently by our MHD simulation, the ambient radiation field must be defined separately. This is comprised of photons contributed from cosmological microwave background (CMB) radiation, and interstellar radiation fields (ISRF) emitted from the host galaxy. We model this as a superposition of four black-body components, with the photon number density given by:
\begin{equation}
    n_{\rm ph}(\epsilon_{\rm ph}, r, z)
  = \sum_i \frac{8\pi\;\! f_{i}(r, z)}{\lambda_{\rm c}^3}
    \frac{\epsilon_{\rm ph}^2}{\exp (\epsilon_{\rm ph}/{\Theta}_i)-1} \ ,   
    \label{eq:isrf_general}
\end{equation}
\citep[e.g.][]{Chakraborty2013ApJ} where the index $i$ denotes the radiation field component (listed in Table~\ref{tab:ISRF_components}), and where $\epsilon_{\rm ph} = h \nu /m_{\rm e} c^2$ is introduced as the dimensionless photon energy (in units of electron rest mass, where $h$ is the Planck constant, $m_{\rm e}$ is the rest mass of an electron and $c$ is the speed of light), $r = \sqrt{x^2+y^2}$ is the radial position of the coordinates $(x, y)$ from the center of the simulation volume at which the radiation field is being evaluated, and $z$ is the height from the $(x,y)$ plane.  
$\lambda_{\rm c}$ is the Compton wavelength of an electron 
  and $\Theta_i = k_{\rm B} T/m_{\rm e}c^2$ is the dimensionless temperature of the radiation field component $i$ (where $k_{\rm B}$ is the Boltzmann constant, and $T_i$ is the temperature of the radiation field component in K). $f_i$ is the intensity factor of the field component $i$ which, for an unmodified black-body radiation field (e.g. the CMB) is set to $f_1 = 1$, but $f_i<1$ in the case of a diluted radiation field (e.g. those which comprise components associated with a galactic ISRF). To model the spatial dependency of the ISRF, we use:
  \begin{equation}
      f_{i}(r, z) = \begin{cases}
    1 \ , \hspace{4.6cm} \text{if} \hspace{0.1cm} i=1\\
    A\;\!\exp(-r/R_{\rm disk}) \;\! \exp(-z/z_{\rm height} ) \ , \hspace{0.5cm} \text{if} \hspace{0.1cm} i>1 
    \end{cases} \ ,
  \end{equation}
  where the component $i = 1$ describes the spatially uniform CMB component. The ISRF spatial distribution roughly follows the exponential disk morphology used to approximate disk-like galaxies (e.g. M31 -- see~\citealt{Courteau2011ApJ}). 
  We adopt $R_{\rm disk} = 10 \;\!{\rm kpc}$ and $z_{\rm height} = 2 \;\!{\rm kpc}$ as the characteristic disk radius and scale height respectively, and set the normalization adjustment $A = \exp(5\;\!{\rm kpc}\;\!/z_{\rm height})$, 
  which yields a broadly similar ISRF extent and intensity to e.g. that available in GALPROP (for details, see~\citealt{Popescu2017MNRAS}). 
The parameter values adopted for the radiation field components are listed in Table~\ref{tab:ISRF_components}, following those suggested for `normal' galaxies by~\citet{Chakraborty2013ApJ} (see also the optical and infrared ISRF contributions in \citealt{Cirelli2009NuPhB}).\footnote{ A similar ISRF model is also adopted by~\citet{Schober2015MNRAS}.} 

\begin{table}
\centering
\begin{tabular}{*{4}{|c|}}
\hline
$i$ & Component & $f_i$ & $T_i/K$ \\
\hline
1 & CMB$^{a}$ & 1 & 2.73 \\
2 & ISRF UV$^{b}$ & $1.6\times 10^{-15}$ & 18000.0 \\
3 & ISRF Optical$^{c}$ & $1.7\times 10^{-11}$ & 3500.0 \\
4 & ISRF IR$^{c}$ & $7.0\times 10^{-5}$ & 41.0 \\
\hline 
\end{tabular}
\caption{Normalization and temperature values adopted for the radiation field model, described by equation~\ref{eq:isrf_general}. This is comprised of the CMB ($i = 1$) and three ISRF ($i = 2, 3, 4$) optical, ultraviolet (UV) and infrared (IR) components. \\
\textbf{Notes}: \\
$^{a}$CMB temperature adopted from~\citet{Planck2018}. \\
$^{b}$UV ISRF component adopted from~\citet{Chakraborty2013ApJ} (see also~\citealt{Schober2015MNRAS}). \\
$^{c}$Optical and IR ISRF components adopted from~\citet{Cirelli2009NuPhB}, according to their Galactic Center region values (also used in the `normal' galaxy models of~\citealt{Chakraborty2013ApJ} and~\citealt{Schober2015MNRAS}). }
\label{tab:ISRF_components}
\end{table} 

\subsection{Bubble development}
\label{sec:bubble_development}

\subsubsection{General characteristics}
\label{sec:gen_char}

\begin{figure*}
\includegraphics[width=\textwidth]{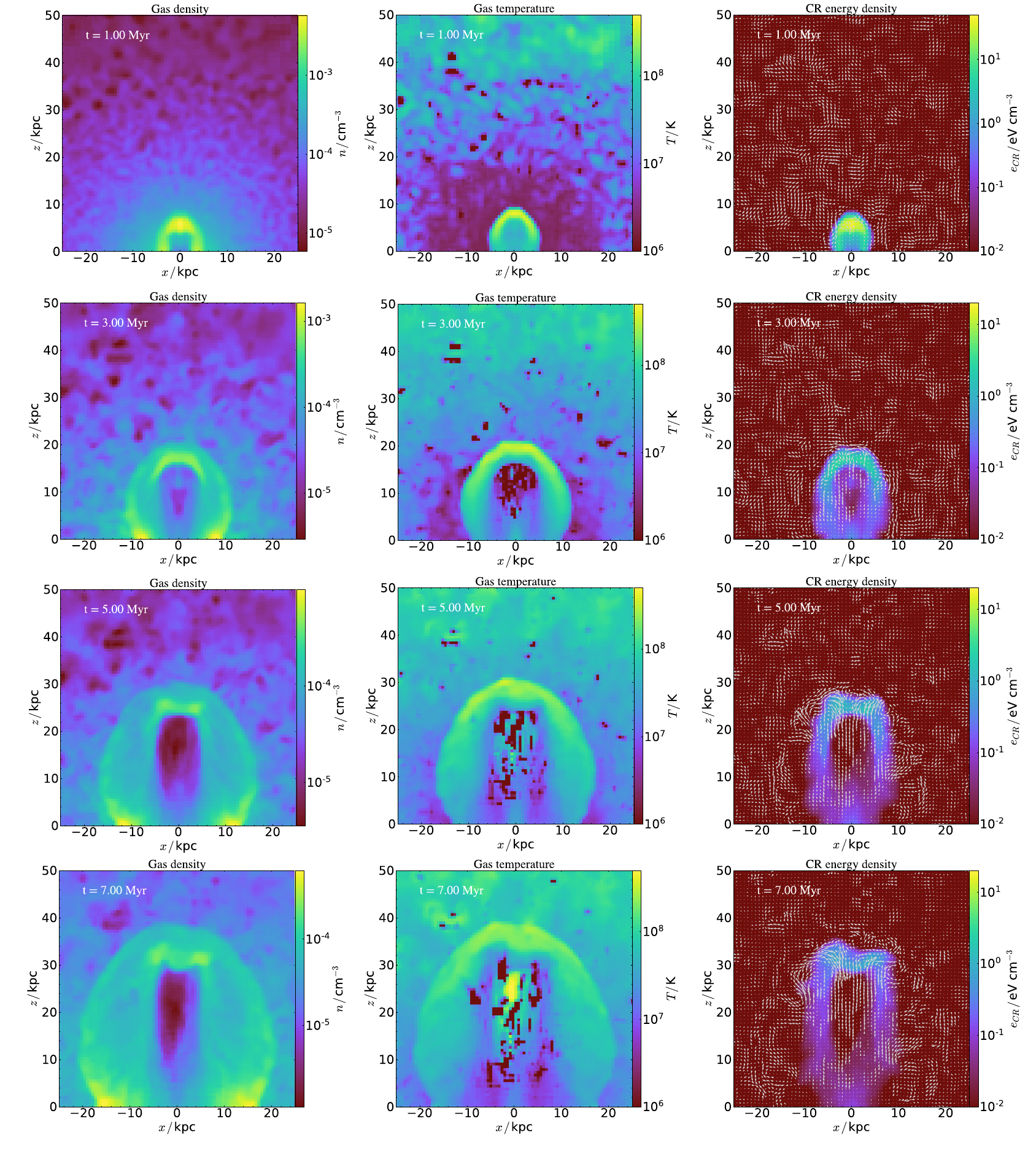}
\caption{2D slice plots through the center of the simulation domain, at $y = 0$, showing the `Northern' region above the galactic plane at $z=0$ (the corresponding bubble below the galactic plane is largely symmetric, with only negligible differences emerging after $\sim$ 2 Myr due to the differences in the magnetic field interfacing the growing bubble). Gas density, temperature and CR energy density, with magnetic field structure over-plotted, are shown at 1 Myr, 3 Myr, 5 Myr and 7 Myr after an injection of CR energy, magnetic field energy and thermal gas is initially provided by bipolar jets at the center of the simulation, and directed along the $\pm z$ directions. {Outside the bubble gas is heated by energy transfer from the background magnetic field, which can be seen to substantially raise the ambient gas temperature by 7 Myr. This effect is artificial and would not arise for a more physically-motivated choice of initial magnetic field model (e.g. one that is weaker away from the host galaxy). However, the impact of this heating is inconsequential to our later results. The dark pixels seen in the temperature plots at 3, 5 and 7 Myr are pockets of cold gas that fall below the temperature scale. }}
\label{fig:bubble_evo}
\end{figure*}

We run our simulations for 7 Myr, which is sufficient to capture the initial growth and subsequent expansion and cooling of the galaxy bubble.  
Figure~\ref{fig:bubble_evo} shows gas density and temperature maps, and CR energy density as a 2D slice through the center of our simulation domain, at $y = 0$ for Run A. Only the 
region above the galactic plane at $z=0$ is plotted, showing the `Northern' bubble. The counterpart bubble below the galactic plane (the `Southern' bubble) is broadly symmetric with respect to $z=0$. Very minor discernible differences between the Northern and Southern bubbles emerge after $\sim$ 2 Myr due to differences in the magnetic field interfacing the expanding structure, but these are inconsequential to our discussion and results. 
In the slice-plot for CR energy density, the local magnetic field orientation and strength (denoted by the direction and size of the over-plotted arrows) is also shown, which co-evolves with the emerging bubble. This shows the case for an initial magnetic field with $\ell_c = 1$ kpc; for comparison, we plot the result for the simulation run with an initial field structure with $\ell_c = 9$ kpc in Figure~\ref{fig:magnetic}. Snapshots are shown at 1 Myr, 3 Myr, 5 Myr and 7 Myr (as labeled) after the start of the simulation, which illustrates the evolutionary stages and development of the system.

\begin{figure*}
\includegraphics[width=0.9\textwidth]{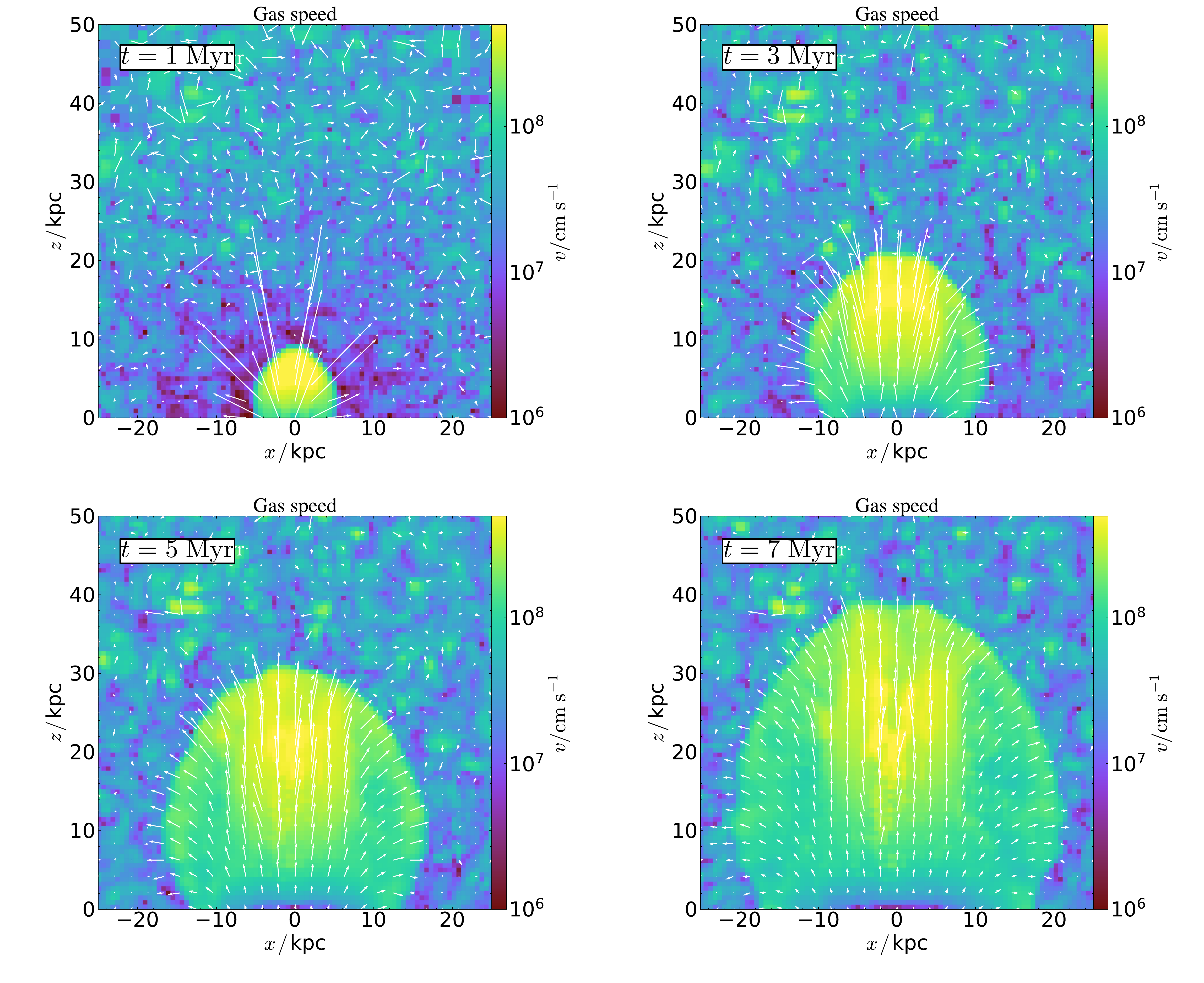}
\caption{2D slice plots through the center of the simulation domain at $y=0$, showing the speed of the hot gas in the developing Northern bubble at 1, 3, 5 and 7 Myr (as labeled) for Run A. The gas velocity is indicated by the over-plotted arrows showing the orientation of the flow, where the size of the arrow reflects the speed. A slowing of the bubble expansion at later times is evident, however the gas within the bubble remains supersonic up to the end of our simulation, at 7 Myr.}
\label{fig:velocities}
\end{figure*}

\begin{figure*}
\includegraphics[width=0.9\textwidth]{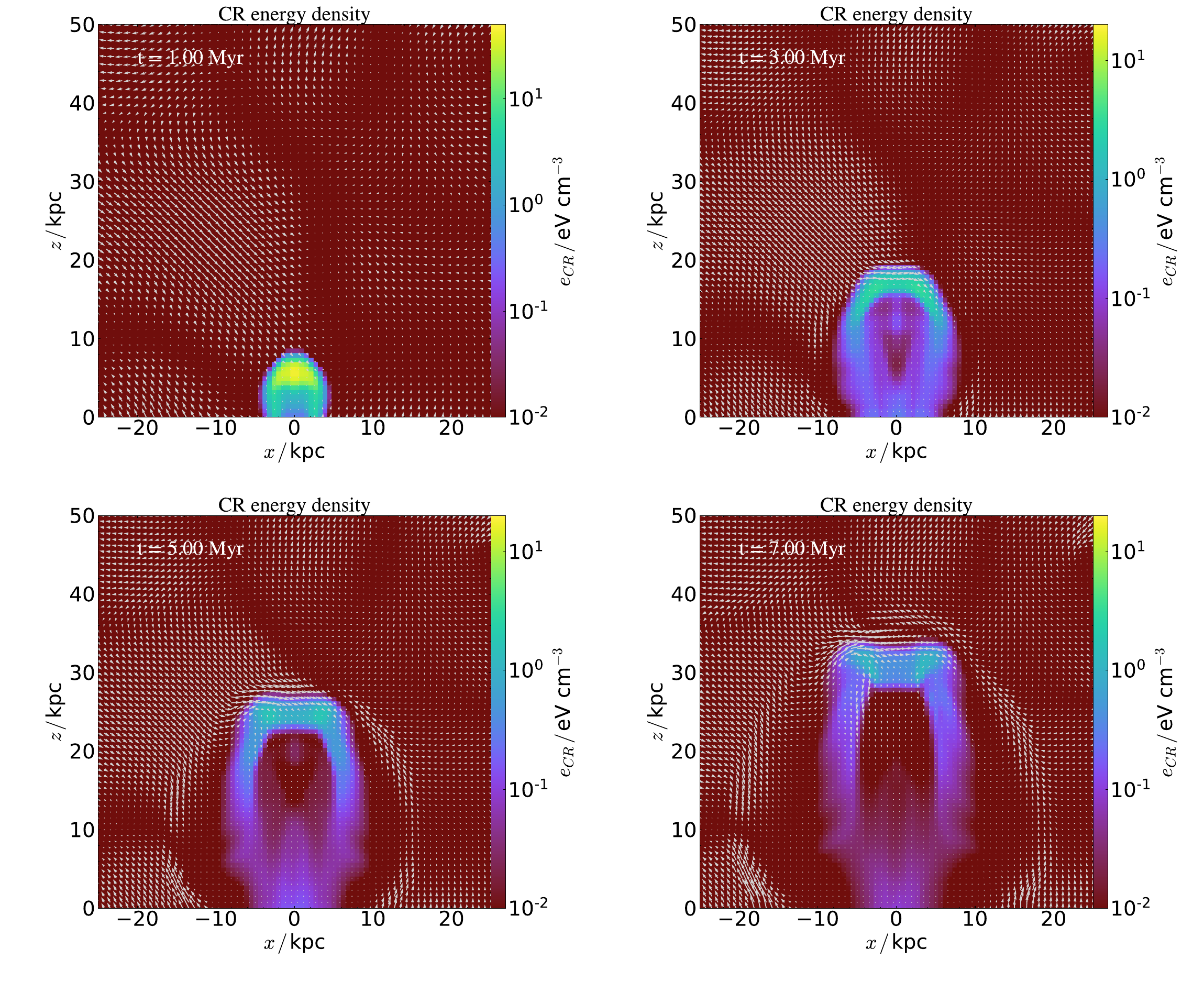}
\caption{2D slice plots through the plane $y=0$ of the simulation plane, showing the CR energy density and overlaid magnetic field vectors for simulation Run B in the Northern bubble at 1, 3, 5 and 7 Myr (as labeled).}
\label{fig:magnetic}
\end{figure*}

An injection of CR energy, magnetic field energy and thermal gas is initially provided by the bipolar jets at the center of the simulation domain and directed along the $\pm z$ directions. These jets are set up to follow the prescription of~\citet{Guo2012ApJ}, although we additionally account for magnetic fields that would be injected by the jets using the AGN sub-grid model of~\citet{Sutter2012MNRAS} using the parameter choices of~\citetalias{Yang2012ApJ}. The jets remain active for the first 0.3 Myr of the simulation, and quickly form the bipolar CR bubbles seen in Figure~\ref{fig:bubble_evo}. {Each bubble is over-pressurized with respect to its ambient medium so, after the jets shut down, the bubble continues to expand under the effects of CR pressure.} CRs are advected by the high velocity of the thermal gas in the jets towards the edges of the bubble (in particular towards the top, following the prevailing velocity of the thermal gas).

The vertical expansion of the CR bubbles in the early stage of their evolution is supersonic, reaching a Mach number of around $M\sim 30$ (with velocities exceeding $10^8 \;\! {\rm cm} \;\! {\rm s}^{-1}$) within the first Myr in most regions in the bubbles, as shown in Figure~\ref{fig:velocities}. 
{The outflow velocities predicted by our model are high, in order to transport the CR electrons before they undergo very substantial cooling. Thus, in our model, the initial jet is kinetically dominated, and the bubble formation time is short. Such high velocity outflows have been inferred by some observational studies towards the Galactic \textit{Fermi} bubbles, using X-ray and UV absorption lines~\citep[e.g.][]{Fox2015ApJ, Miller2016ApJ, Bordoloi2017ApJ}, though there remain uncertainties in the modeling and interpretation of the data~\citep[for discussion, see][]{Yang2018Galax}.} 

Ambient hot halo gas is compressed by the resulting forward shocks into a shell around the bubbles, where gas temperatures exceed 10$^8$ K and densities reach around 10$^{-2}$ cm$^{-3}$. This is separated by a contact discontinuity from the $T\geq 10^{7}$ K, $n\sim 10^{-4} \;\! {\rm cm}^{-3}$ gas inside the bubbles. The expansion of the bubbles in the lateral direction, perpendicular to the driving direction of the initial jets, is somewhat slower than in the vertical direction, with Figure~\ref{fig:velocities} indicating velocities of around $10^7 -10^8 {\rm cm}\;\! {\rm s}^{-1}$ being attained (corresponding to Mach numbers of around $M \sim 10-12$), and temperatures and densities seeing less enhancement inside the lateral shocks ($\sim 10^{7}$ K and $\sim 10^{-3}$ cm$^{-3}$, respectively) than in the vertical shocks. By 1 Myr, a significant concentration of CR energy has developed in the bubbles above $\sim$ 5 kpc, reflecting the role CRs play in sustaining the rapid expansion of the bubble front. An excess of CR energy density can also be seen to be developing laterally. This contributes to the broadening of the bubbles together with the thermal gas pressure, with the observed convex morphology arising from a decline in the inward pressure imparted by the ambient halo gas at increasing altitudes. The development of a large low-density central lobe ($n\sim 10^{-5}$ cm$^{-3}$) within the bubbles is also evident after 1 Myr, enclosed by an inner contact discontinuity. This continues to expand with the bubble, and contains very high temperature jet plasma and jet-entrained halo gas. Very high temperatures are sustained within these central lobes throughout the evolution of the bubbles, and still exceeds $10^8$ K by $\sim 7$ Myr in some regions. Surrounding these lobes, between the inner contact discontinuity and the forward shock, outer shells containing shock-heated halo gas can be seen. The magnetic field in these outer shell regions is relatively ordered and is typically oriented vertically, aligned with the direction of the initial jets.

The magnetic field of the bubbles, shown in the right panels of Figure~\ref{fig:bubble_evo}, co-evolves with the gas.\footnote{{We note that our simulations also show an evolution of gas temperature and density structure {of the ambient gas} \textit{outside} the bubble. We find this arises because the tangled initial magnetic field acts to perturb the gas distribution on scales comparable to its correlation length. The dissipation of these perturbations leads to a heating effect on the external thermal gas, with the magnetic energy density in the regions outside the bubble reducing at a similar rate to the increase in the thermal energy of the gas. This process can be considered artificial and physically overstated, and does not arise if a more } {physically-motivated initial magnetic field configuration is adopted (e.g. one that is weaker at greater distances from the host galaxy).}} As the bubbles expand, a magnetic draping effect ~\citep{Lyutikov2006MNRAS} is evident around their surface, where magnetic field lines are stretched and compressed into a draping layer surrounding the bubbles. This process 
aligns the magnetic field to lie parallel to the {bubble's outer shocked layer}, and can strengthen the field as it is compressed against the expanding bubble. 
The strengthening and aligning effect by draping is greater when the direction of the field is initially parallel to a bubble surface, with the field strengths sometimes reaching more than 10 times their initial value in the draping layer by 1 Myr. This has important consequences for CR propagation: while the CRs are advected to the surface region of the bubbles, anisotropic diffusion governs their distribution near and within the draping layer. With CR diffusion being directed along the local magnetic field vector, the ordered magnetic field in the draping layer effectively contains the CRs, creating a sharp boundary in CR energy density at the leading edge of the bubbles. This helps to maintain CR pressure in this region, and contributes to bubble expansion.

This sharp boundary in CR energy density begins to diminish from around 3 Myr, particularly at lower altitudes where the lateral shock continues to propagate into the ambient medium, but where the CRs are not as effectively advected to the {leading shock layer}. In this region, 
the lower outward gas velocities away from the expansion axis of the bubbles (cf. Figure~\ref{fig:velocities}) increase CR advection times, allowing diffusive propagation to have more influence on the distribution of CRs. 
As expansion continues, the lateral shock front moves increasingly further away from the bulk of the CR energy density, leaving a disordered magnetic field in its wake. The gas in the CR-deficient layer behind the lateral front cools adiabatically, reducing to $\sim 3\times 10^{7}$ K by 5 Myr (below $\sim$10 kpc). The resulting reduction in lateral CR and thermal pressure slows the broadening of the bubbles, particularly after 3 Myr (the point at which the lateral shock decouples from the CR layer), leading to the emergence of much more elongated bubbles by 7 Myr. The magnetic draping effect also begins to weaken after 3 Myr, and by 5 Myr is only clear in the upper regions of the bubble where CR energy density remains high and connected to the {leading shock}. By 7 Myr, the draping effect is only evident in small regions of the bubble (e.g. the North-Western corner) where CR energy densities remain high and bubble expansion is most rapid (cf. bottom-right panel of Figure~\ref{fig:velocities}). 

\subsubsection{Effect of the initial magnetic field}
\label{sec:mag_init_effect}

\begin{figure*}
\includegraphics[width=\textwidth]{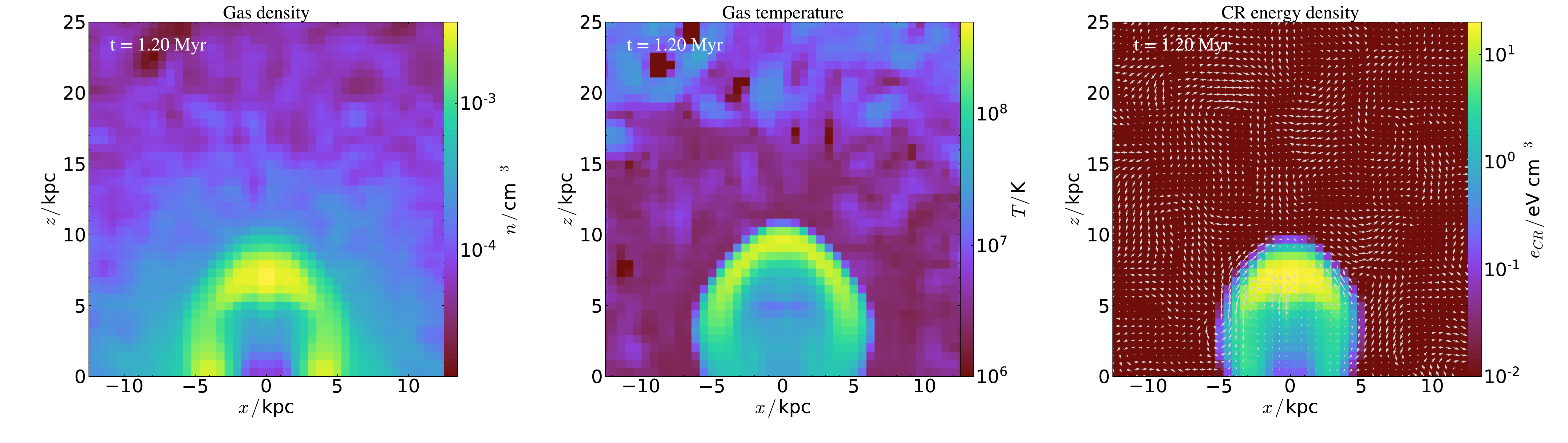}
\caption{2D slice plots of density, temperature and CR energy density for Run A at 1.2 Myr, which can be compared to the previous results of~\citetalias{Yang2012ApJ}. Note that the region plotted is smaller than that shown in earlier Figures.}
\label{fig:bubble_1p2_compare}
\end{figure*} 

We find that the configuration of the initial halo magnetic field can have noticeable implications for the late-time morphology of the bubbles. Figure~\ref{fig:magnetic} shows the evolution of the CR energy density and magnetic field for our simulation Run B, where we initialize the halo magnetic field with a correlation length of $\ell_c = 9$ kpc, instead of $\ell_c = 1$ kpc (as was used for the Run A results shown previously in Figure~\ref{fig:bubble_evo}). The simulations are otherwise initialized identically. 
 When comparing Figure~\ref{fig:magnetic} (Run B) to the right-hand panels of Figure~\ref{fig:bubble_evo} (Run A), small-scale features can be seen to arise at the bubble edges in Run A, when the initial magnetic field has a correlation length that is much smaller than the size of the bubbles. These are not present in Run B. This is due to fast CR diffusion along field lines, and the varying effect of magnetic draping over the {expanding shock layer} of the bubble in Run A. In Run B, when the correlation length of the magnetic field is comparable to the bubble scale (Figure~\ref{fig:magnetic}), a more even surface results due to less randomly oriented field directions on small scales. It can also be seen that magnetic draping persists around the {outer} shock for much longer, and still presents clearly after 7 Myr (bottom right panel, Figure~\ref{fig:magnetic}). This maintains a smoother surface in the bubbles.
 Although our initial magnetic field is unlikely to be the same as a more physically-motivated model, our results imply that the exact choice of initial field structure is more likely to impact 
{small-scale} features, account for deviations from absolute symmetry in emerging bubbles, or lead to distortions of the surface of the bubbles (we note that \citetalias{Yang2012ApJ} drew similar conclusions in the context of simulations of the Milky Way's \textit{Fermi} bubbles). The overall evolution and general features of the bubbles are not significantly affected by the initial magnetic field topology and, for the purposes of this work, the kpc-scale distribution of CRs, gas density and temperature is of most importance -- none of which would be impacted dramatically {by} plausible variations in the initial magnetic field.

\subsubsection{Comparison with previous work}
\label{sec:mhd_comparison_lit}

Recent theoretical studies have considered the development of CR bubbles in the context of the Milky Way's own \textit{Fermi} bubbles. As possible examples of jet-driven CR bubbles~\citep[e.g.][]{Guo2012ApJ_b, Guo2012ApJ, Yang2012ApJ, Yang2017ApJ, Zhang2020ApJ, Zubovas2011MNRAS}, these offer a wealth of results suitable for comparison with the earlier stages of the CR bubble simulated in this work. The closest comparison can be made with~\citetalias{Yang2012ApJ}, upon which the simulations of the present study are based. The evolution of the bubbles are tracked until 1.2 Myr, at which time the~\citetalias{Yang2012ApJ} results were compared to available constraints and observed properties of the \textit{Fermi} bubbles. Although the evolution to later times was not previously considered, our simulations at 1.2 Myr can still be reasonably compared to the earlier work.

Figure~\ref{fig:bubble_1p2_compare} shows density, temperature and CR energy density 2D slice plots of our simulations at 1.2 Myr, which are directly comparable to the results at 1.2 Myr presented in~\citetalias{Yang2012ApJ} (their Figure 1). Overall, the structure and properties of this work is broadly consistent with the earlier result, including magnetic draping effects, the bubble morphology and the fast inflation timescale, within a few Myrs. However, certain differences are evident - in particular, 
our simulations give a more elongated bubble; while the broadening to a similar maximum radius of $\sim 5$ kpc is reached at a comparable altitude of $z\sim 3$ kpc in both cases, the peak of the bubble and distribution of CRs is sharper in this work than in~\citetalias{Yang2012ApJ}. Moreover, a dent feature resulting from the cuspy initial gas density profile\footnote{The density gradient across the width of the initial jets is not negligible, thus it is more difficult for the central parts of the jets to penetrate the dense material near the center of the host galaxy. If a density profile model were adopted with a flatter core, the resulting bubble would expand slightly faster in the $\pm z$ direction, and would show a flatter top.}
is well-resolved
in the 1.2 Myr results of~\citetalias{Yang2012ApJ}, but does not clearly emerge in the present work until nearly 3 Myr. Because of this, the top of the bubble in~\citetalias{Yang2012ApJ} extends to $\sim 7$ kpc, much less than the $\sim$10 kpc extent seen in Figure~\ref{fig:bubble_1p2_compare}. By running higher resolution simulations for shorter times, we were able to attribute these differences to the lower resolution set-up adopted in the present work. Our lower resolution choice was necessary due to the substantially larger simulation volume (box side length 75 kpc here compared to 25 kpc in~\citetalias{Yang2012ApJ}) and longer simulation time (7 Myr here, compared to 1.2 Myr previously) required for the focus of this project, in the context of computational constraints. We anticipate future studies will be able to run simulations at comparable resolution to~\citetalias{Yang2012ApJ}, with larger volumes and for longer times, however we do not expect the lower resolutions of this work to substantially change our results.

\cite{Yang2017ApJ} extended the treatment of~\citetalias{Yang2012ApJ}, by also tracking the evolution of the CR electron spectrum during their simulations. However, in this updated treatment CR transport was modeled as advection, with diffusion omitted. This led to the emergence of a sharper bubble edge than seen either in~\citetalias{Yang2012ApJ} or this work, where anisotropic diffusion is included. \cite{Yang2017ApJ} additionally used a smoothed halo potential, suppressing the formation of the dent feature at the top of the bubbles.

Comparison may also be made with the results of~\citet{Guo2012ApJ}, which also demonstrated that galactic-scale CR bubbles could be inflated rapidly to kpc scales, within a few Myr, by a short-duration ($< 0.5$ Myr) energetic AGN outburst. While, again, their bubble evolution is not dissimilar to our simulations (cf. the first two rows of panels in Figure~\ref{fig:bubble_evo}) and also shows the development of a shock which heats and compresses ambient gas in the halo, a crucial difference is the appearance of strong instabilities in the~\citet{Guo2012ApJ} results that are not present in Figure~\ref{fig:bubble_1p2_compare} (or, indeed Figure~\ref{fig:bubble_evo} up to 7 Myr). \citetalias{Yang2012ApJ} previously showed that these instabilities were a consequence of the longer estimated bubble formation time in~\citet{Guo2012ApJ}, which were sufficient to allow the development of Kelvin-Helmholtz (KH) instabilities (over a timescale $\tau_{\rm KH}$ of around 1.5 Myr). \citetalias{Yang2012ApJ} further discussed that such instabilities would emerge if their simulations were allowed to run for longer, i.e. for $t_{\rm sim} > \tau_{\rm KH}$. 
However, when accounting for the evolution of the bubble size and shear velocity (considering only the vertical velocity component with {$\Delta v_z\sim 100\;\!{\rm km s}^{-1}$} at the outer-edge of the bubble, e.g. {on the contact discontinuity} at a position of {$x=10 \;\! {\rm kpc}, y=0 \;\! {\rm kpc}, z=20 \;\! {\rm kpc}$} at 7 Myr in {Figure~\ref{fig:velocities}}) the KH instability timescale increases faster than the ageing of the bubble, reaching 
\begin{equation}
    \tau_{\rm KH} \approx 100 \; \left( \frac{\lambda_{\rm s}}{20\;\!{\rm kpc}}\right) \left( \frac{\Delta v_{\rm s}}{100 \;\!{\rm km}\;\!{\rm s}^{-1}} \right)^{-1} \left( \frac{\eta_{\rm s}}{0.1} \right)^{-1/2} ~{\rm Myr}
\end{equation}
by 7 Myr, where
$\lambda_{\rm s}$ is the wavelength of the instability (we consider modes comparable to the radius of the bubble which would have significant effects on the bubble's macroscopic structure), 
$\Delta v_{\rm s}$ is the shear velocity at the surface of the bubble\footnote{From Figure~\ref{fig:velocities}, it can be seen from the over-plotted velocity arrows that the outward $\pm x$ gas velocity far dominates over the shear velocity $\pm z$ at this position. We find 
that $v_z\sim 0.1 |v|$, where $|v|$ is the overall magnitude of the gas velocity (as plotted in Figure~\ref{fig:velocities}).}, 
and $\eta_{\rm s}$ is the density contrast across the contact discontinuity at the bubble surface.
This suggests that the instabilities seen in~\citet{Guo2012ApJ} would not be expected at \textit{any} point during the lifetime of a bubble in our simulations. 

The approach adopted more recently by~\cite{Zhang2020ApJ} is broadly the same, again with the focus of modeling the Galactic \textit{Fermi} bubbles. They consider the development of a CR bubble due to an AGN jet, operating with less energy ($\sim 10^{55}$ erg) compared to our work (which used a total injected energy of $\sim$ 10$^{57}$ erg per jet) but over a longer period of time (1-5 Myr, compared to 0.3 Myr here), with CR and thermal energy injected by the jet. Their high resolution simulations allow a clear view of the development of a bubble comprising of a low-density central lobe, which is enclosed by a contact discontinuity and containing high-temperature jet plasma and some jet-entrained halo gas. An outer shell located between a forward shock and inner contact discontinuity containing the shock-heated halo gas is also shown to form.\footnote{{Note that, in contrast to our work,~\citet{Zhang2020ApJ} define the surface of the bubbles by the forward shock instead of the contact discontinuity. Either definition is reasonable, and we find KH instabilities of wavelengths comparable to the radius of the bubble would not develop in our model if adopting either definition.}} These features are common to our work, and the resulting bubble structure is the same. However, the alternative lower choice of jet power and longer activity timescales lead to a longer bubble expansion time, taking around 5 Myr to extend to 10 kpc (for a jet energy of $1.07 \times 10^{55}$ erg and activity time of 1 Myr), compared to $\sim 1$ Myr in our work. We consider that the~\cite{Zhang2020ApJ} results give some insights into the behavior of our own model with alternative jet parameter choices, and illustrates the variation in plausible bubbles that may form around galaxies. 

\section{Multi-wavelength emission from CR electrons in galaxy bubbles}
\label{sec:section3}

Emission thought to originate from the \textit{Fermi} bubbles of our Galaxy has been observed over a broad range of wavelengths, from radio waves~\citep{Heywood2019Natur} and
microwaves~\citep{Finkbeiner2004ApJ, Planck2013AA, Sasaki2015ApJ}, through to 
X-rays~\citep{Ponti2019Natur, Predehl2020Natur}
and $\gamma$-rays~\citep{Su2010ApJ, Dobler2010ApJ} up to $\sim 110$ GeV~\citep{Ackermann2014ApJ} -- and, in some regions, perhaps reaching energies as high as 500 GeV~\citep{Herold2019AA}. In a leptonic scenario, this emission is predominantly generated by a population of CR
 electrons within the bubbles through direct synchrotron, Compton scattering and bremsstrahlung processes. The internal magnetic, density and thermal radiation structure of the bubbles determines the emission spectrum and topology that arises thus, for the \textit{Fermi} bubbles and analogous leptonic CR bubbles around external galaxies, spectral and (where possible) spatial studies of this emission can yield detailed information about the physical conditions in different parts of the bubble. 

Our simulations (Figure~\ref{fig:bubble_evo}) have shown that the conditions within a developing CR bubble evolve over Myr timescales. Substantial variations in the magnetic field strength and structure, gas temperature and density can be seen as a bubble ages and expands. Moreover, the internal distribution of CRs shows a similar degree of variation over a bubble lifetime. As the emitted multi-wavelength spectrum of a leptonic CR bubble would be sensitive to the MHD properties, as well as the spatial distribution of CR electrons\footnote{The spectrum of electrons throughout a bubble is likely to be relatively uniform until late times, owing to their fast expansion~\citep{Yang2012ApJ, Yang2017ApJ}.}, it would retain crucial information necessary to probe their internal MHD properties, composition and the evolutionary stage. 

Full, detailed emission maps are likely only within observational reach for the Milky Way's \textit{Fermi} bubbles over a very large range of wavelengths. However, CR bubbles around external, distant, galaxies may still be observed spectrally in one or more wavebands, allowing for the extraction of information about their internal conditions. Even for very distant systems, for which the number of arriving photons may be small, carefully selected observation bands may still be sufficient to provide constraints on the physical conditions inside an emitting galaxy bubble, avoiding the need for the long integration times or high sensitivities that would be required for the construction of a full spectrum. Modeling expected spectral signatures for bubbles in different evolutionary stages can therefore offer information about the most valuable spectral signatures and, hence, the most appropriate energy bands to search for to extract certain information about bubbles around distant galaxies.
This can increase the useful reach of observational studies and boost the number of observable galaxy bubbles, offering the potential to gather broader data to inform detailed population analyses. Such analyses will be crucial to understand the nature and evolution of CR galaxy bubbles, and how our own Galaxy's \textit{Fermi} bubbles fit into a larger galaxy bubble demographic.

\subsection{CR electron interactions}
\label{sec:electron_interactions}

CR electron interactions can broadly be modeled as cooling processes -- i.e. either 
processes in which the electrons truly experience a continuous loss of energy, or stochastic processes where the energy lost by the CR electron in a single interaction/scattering event is only a small fraction of its total. 
The radiative interactions of CR electrons in a galaxy bubble would predominantly drive the observable emission in the leptonic scenario. 
The relevant processes are synchrotron losses in the ambient magnetic fields, and Compton scattering in ambient radiation fields. The energy loss rate for an electron undergoing synchrotron cooling is given by:
\begin{equation}
 \frac{{\rm d}\gamma_{\rm e}}{{\rm d}\;\!t}\;\!\bigg\vert_{\rm sy} 
    = \frac{4}{9} \left[\frac{e^4B^2c}{(m_{\rm e}c^2)^3}\right] \gamma{\rm e}^2 \beta^2
  \approx \frac{32}{9} 
  \left(\frac{2 \sigma_{\rm T}}{m_{\rm e} \;\!c}\right) 
       U_{\rm B} \gamma_{\rm e}^2     \ , 
         \label{eq:synch_cooling}
\end{equation}
where 
   $\beta$ is introduced as the speed of the particle, normalised to the speed of light, $c$,  
   $\sigma_{\rm T}$ is the electron Thomson cross-section, 
  $e$ is electron charge, 
  $B$ is the magnetic field strength, and 
  $U_{\rm B}(\equiv B^2/8\pi)$ is the magnetic energy density.  
Other symbols retain their earlier definitions. 
 For (inverse) Compton cooling, the corresponding rate of energy loss for a relativistic electron is: 
\begin{equation}
 \frac{{\rm d}\gamma_{\rm e}}{{\rm d}\;\!t}\;\!\bigg\vert_{\rm ic} 
     \approx \frac{32}{9} 
     \left(\frac{2 \sigma_{\rm T}}{m_{\rm e}\;\!c}\right)  
            U_{\rm rad} \gamma_{\rm e}^2  \ , 
         \label{eq:compton_cooling}
\end{equation}
  which depends mostly on the energy density of the ambient radiation field, 
  i.e. $U_{\rm rad}$. This would be dominated by 
  the energy density of the CMB radiation in the conditions expected for a galaxy bubble and the ISRF at low altitudes near the host galaxy (see section~\ref{sec:gal_rad_field}).
  
  Electrons may also experience losses due to their engagement with the ambient hot plasma within the bubble. 
  Bremsstrahlung (free-free) losses proceed at a rate given by:
\begin{equation}
\frac{{\rm d}\gamma_{\rm e}}{{\rm d}\;\!t}\;\!\bigg\vert_{\rm ff} \approx \alpha_{\rm f} {c} \sigma_{\rm T} n_{\rm H} \gamma_{\rm e} \,
\label{eq:ff_cool_elec}
\end{equation}
where $\alpha_{\rm f}$ is introduced as the fine structure constant, and $n_{\rm H}$ is the plasma number density.
Also Coulomb (Rutherford) scattering losses are given by 
\begin{equation}
\frac{{\rm d}\gamma_{\rm e}}{{\rm d}\;\!t}\;\!\bigg\vert_{\rm C} \approx n_{\rm H} {c} \sigma_{\rm T} \ln \Lambda \ ,
\label{eq:c_cool_elec}
\end{equation}
where $\ln \Lambda\simeq 30$ is the Coulomb logarithm, accounting for the ratio between the maximum and minimum impact parameters. Adiabatic losses are also relevant in the rapidly expanding high-energy bubbled modeled in this work. These arise at a rate given by:
\begin{equation}
    \frac{{\rm d}\gamma_{\rm e}}{{\rm d}\;\!t}\;\!\bigg\vert_{\rm adi} = \frac{\gamma_{\rm e}}{3}\;\!\frac{{\rm d}{\rm ln} V}{{\rm d}t} \ ,
    \label{eq:c_cool_adi}
\end{equation}
where $V$ is the bubble volume. The total cooling rate experienced by a CR electron would be the sum of all of these processes (equations~\ref{eq:synch_cooling} --~\ref{eq:c_cool_adi}).

\subsection{Electron spectrum}
\label{sec:leptonic_bubbles}

\begin{figure}
\includegraphics[width=\columnwidth]{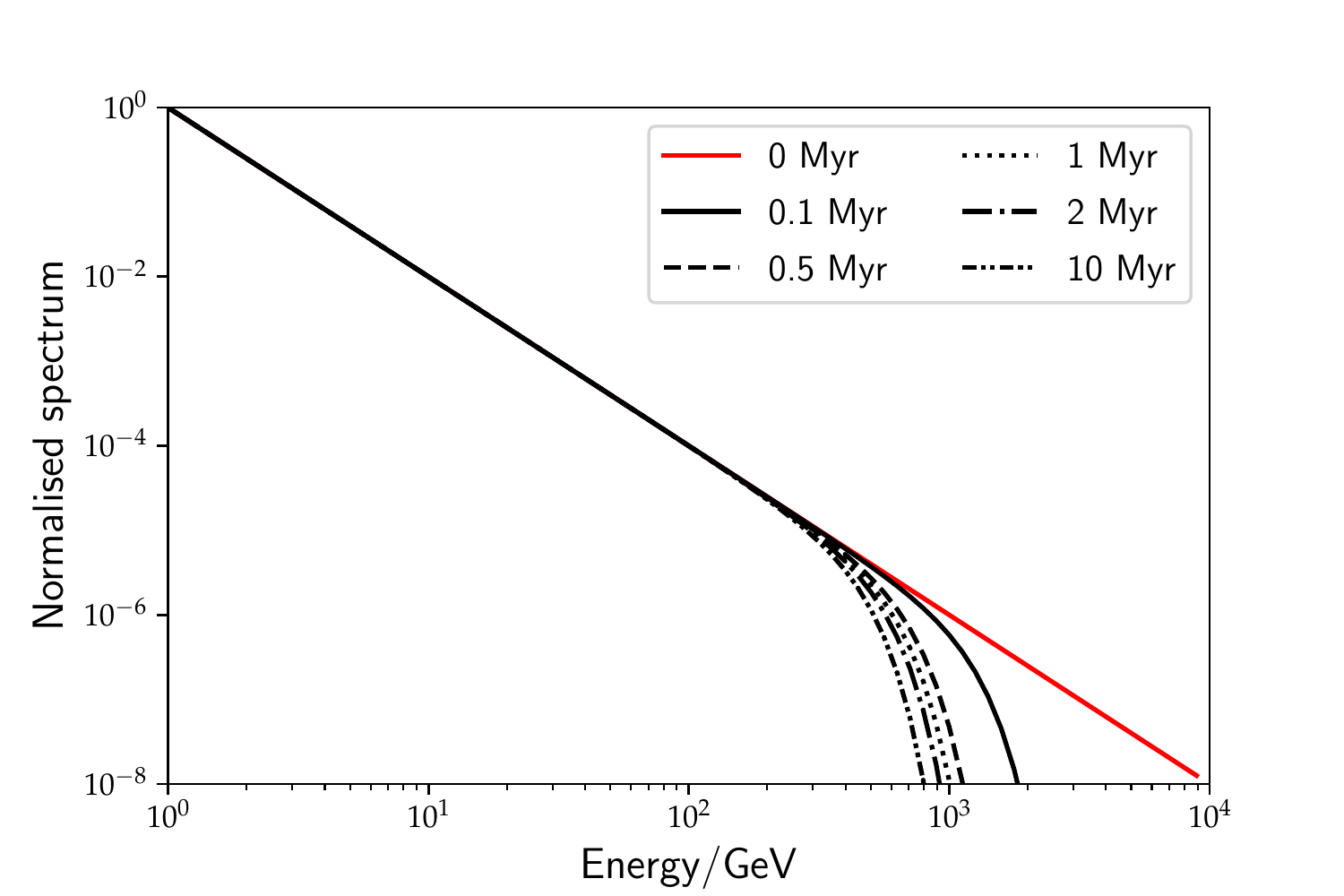}
\caption{Evolution of the CR electron spectrum, according to our analytic prescription (see Appendix~\ref{sec:cr_spec_evo} for details). The initial power-law spectrum (given by equation~\ref{eq:electron_spectrum}) is shown in red, and the subsequent spectrum at 0.1, 0.5, 1, 2 and 10 Myr is shown as labeled. This illustrates the very rapid initial ageing of the spectrum at high energies, followed by very little change after $\sim$0.5 Myr.} 
\label{fig:spec_ageing}
\end{figure}

The emitted radiation from a CR bubble depends on the CR spectrum. Typically, assumptions about the spectrum must therefore be made to make observable predictions (unless the CR spectral evolution is tracked directly in a model or simulation, as in~\citealt{Yang2017ApJ}). The most simple reasonable model that is frequently assumed is a power law, with distinct upper and lower spectral cuts~\citep{Su2010ApJ, Yang2013MNRAS}. 
In~\citealt{Yang2017ApJ}, it was demonstrated that the fast electron cooling that happens close to the GC leads to a relatively uniform spectrum with a 1000 GeV cut-off throughout the whole bubble during the early expansion phase of a bubble, and this result is not strongly sensitive to the exact time {the CRs are} injected. However, the precise form of the spectrum they found was not a clear power-law at all energies, and showed steepening above a few hundred GeV. 

In this work, we do not track the evolution of {the} CR spectrum during our simulations. However, to emulate the spectral evolution found in~\citealt{Yang2017ApJ}, we adopt an \textit{initial} power law CR electron differential energy spectrum between $E_{\rm min} = 1$ GeV\footnote{We consider that lower energy electrons, below 1 GeV, would rapidly lose energy via Coulomb collisions in the early stages of the bubble development, within the first 0.1 Myr, due to the high gas density in the central regions of the host galaxy from where the bubble is launched. Our choice would correspond to a gas density of around 500~${\rm cm}^{-3}$ initially encountered by the electrons, which is broadly consistent with minimum mean gas densities estimated for the Milky Way galactic center region (of around 100~${\rm cm}^{-3}$ -- see~\citealt{Oka2005ApJ}), conservative compared to typical estimates ($\sim 10^4\;\!{\rm cm}^{-3}$ -- see, e.g.~\citealt{Guesten1983AA, Bally1987ApJS, Mills2018ApJ}), and consistent with the density ranges inferred for the central regions of other nearby galaxies (e.g. M74 -- see~\citealt{Heiner2013MNRAS}).} and $E_{\rm max} = 10,000$ GeV, which we then evolve analytically to give the electron spectrum at some time $t$, accounting for cooling losses in the early stage of the bubble expansion. The initial spectral form may be written as:
\begin{equation}
\frac{{\rm d}n_{\rm e}}{{\rm d}\gamma_{\rm e}} = \mathcal{N}_e \left(\frac{\gamma_{\rm e}}{\gamma_0}\right)^{-s_{\rm e}} 
\label{eq:electron_spectrum}
\end{equation}
where 
$n_{\rm e}$ is the CR number density, 
$\gamma_{\rm e}$ is the electron Lorentz factor, (which is related to the electron energy by $E_{\rm e} = \gamma_{\rm e} \;\! m_{\rm e} c^2$), 
$\gamma_0 = E_{\rm min}/m_{\rm e} c^2$ is the parameter chosen to set the lower energy cut-off of the CR electron spectrum,
$s_{\rm e} = 2$ is the power-law spectral index. In general, 
the normalization $\mathcal{N}_e$ is set by the CR energy density $e_{\rm CR}$ in a simulation cell, according to the relation 
\begin{equation}
\mathcal{N}_e = \begin{cases}
\frac{f_{\rm emit}\;\!e_{\rm CR} \gamma_0^{-s_{\rm e}} }{\ln (\gamma_1/\gamma_0)} \hspace{2.9cm} \text{if}\;\!s_{\rm e} = 2 \ , \\
\frac{f_{\rm emit}\;\!e_{\rm CR} (2-s_{\rm e}) \gamma_0^{s_{\rm e}}}{\gamma_{1}^{2-s_{\rm e}} - \gamma_{0}^{2-s_{\rm e}}} \hspace{2.3cm} \text{if}\;\!s_{\rm e} \neq 2 \,
\end{cases}
\label{eq:electron_norm}
\end{equation}
which introduces $\gamma_{1} = E_{\rm max}/m_{\rm e} c^2$ as the upper spectral cut-off. {$f_{\rm emit}$ retains its earlier definition and value (see section~\ref{sec:gal_rad_field}).}

To model the ageing of the CR electron spectrum due to cooling processes, we solve the appropriate kinetic equation governing the spectra of CRs (e.g. ~\citealt{Kardashev1962SvA} -- see Appendix~\ref{sec:cr_spec_evo} for details), where we neglect acceleration processes, particle absorption, injection or escape, and assume cooling is due to radiative losses (synchrotron cooling in ambient magnetic fields, and inverse Compton cooling in the CMB, and the ISRF near the host galaxy) and adiabatic losses. We adopt equation~\ref{eq:electron_spectrum} as the initial condition. To approximate the adiabatic and radiative cooling rate of the electrons in a manner which reflects the evolution of our simulated system, we estimate electron cooling rates from the timescales determined from tracer particles in~\citet{Yang2017ApJ}, which self-consistently tracked the electron spectral evolution within their simulation over the first 1.2 Myr.\footnote{We consider this to be a reasonable approach, given the practical similarities between the simulations in this work and those of~\citet{Yang2017ApJ}.} Cooling rates at later times were then extrapolated as required. 

The resulting spectral ageing is shown in Figure~\ref{fig:spec_ageing}, where the initial spectrum is shown together with the aged spectrum at various stages during the evolution of the bubble. This reflects the rapid initial cooling and otherwise marginal evolution seen in~\cite{Yang2017ApJ}, and also would reproduce a relatively uniform spectral shape throughout the bubble. A natural cut-off can be seen to emerge at around 1000 GeV, and the resulting spectrum is very similar in form to that assumed in the post-processing calculations of~\cite{Yang2013MNRAS}. It is also evident that the effects of adiabatic cooling on the electron spectrum is negligible (this would affect all electrons independent of their energy and, if significant, would be seen as a lowering of the spectrum over time). For our calculations, we assume the same spectral form throughout our simulation, specified at the required time, but vary its normalization according to the local simulated CR energy density $e_{\rm CR}$ (cf. equation~\ref{eq:electron_norm}).

\subsection{Multi-wavelength emission}
\label{sec:emitted_spectrum}

The spectral emissivity from CR electrons is given by the sum of   
the emission from relevant radiative processes (section~\ref{sec:electron_interactions}):
\begin{align}
j(\epsilon) & = \frac{{\rm d}\epsilon}{{\rm d}t\;\!{\rm d}\nu\;\!{\rm d}V}  \nonumber \\
&= j_{\rm sy}(\epsilon) + j_{\rm ic}(\epsilon) +j_{\rm ff}^{\rm nt}(\epsilon)+j_{\rm ff}^{\rm t}(\epsilon) \ ,
\end{align}
 i.e. the sum of synchrotron, inverse Compton, (non-thermal and thermal) bremsstrahlung spectral emissivities. 
The synchrotron spectral emissivity for a population of CR electrons with differential number density ${{\rm d} n_{\rm e}}/{{\rm d}\gamma_{\rm e}}$ is then given by:
\begin{align}
 j_{\rm sy}(\epsilon) & = \frac{4\sqrt{3} \pi r_{\rm e}}{9 \;\! c} \frac{\nu^2}{\nu_{\rm B}} \;\! \int_{\gamma_0}^{\gamma_1} \frac{{\rm d} \gamma_{\rm e}}{\gamma_{\rm e}^{4}} \frac{{\rm d} n_{\rm e}}{{\rm d}\gamma_{\rm e}} \;\! \mathcal{F}(x)
 \label{eq:sy_emission}
\end{align}
\citep{Dermer2009_book}, where ${\rm d}V$ is the differential volume element, 
$\nu$ is the emitted spectral frequency, 
$\epsilon = h\nu/m_{\rm e} c^2$ is the dimensionless energy of the emitted photons in units of electron rest mass,
$r_{\rm e}$ is the classical electron radius, 
$\nu_{\rm B} = e B/(2\pi m_{\rm e} c)$ is the electron gyro-frequency, 
and 
$x = \nu/(3 \nu_{\rm B} \gamma_{\rm e}^2)$. The function  $\mathcal{F}(x)$ is given by:
\begin{equation}
\mathcal{F}(x) = \left( K_{4/3}(x) \;\! K_{1/3}(x) - \frac{3}{5} x \left[ K_{4/3}^2(x) - K_{1/3}^2(x)\right] \right) \ ,
\end{equation}
where $K_{m}(\hdots)$ are modified Bessel functions of the second kind, of fractional order $m$.
CR electrons may also radiate by Compton scattering off ambient low-energy (CMB and, near the host galaxy, ISRF) photons. 
For a radiation field of photon density $n_{\rm ph}$, the inverse Compton emissivity is given by: 
\begin{equation}
 j_{\rm ic}(\epsilon) = \frac{3 \sigma_{\rm T} \lambda_c^2 m_{\rm e} c \nu}{4} \int_{\gamma_{0}}^{\gamma_{1}} \frac{{\rm d}\gamma_{\rm e}}{\gamma_{\rm e}^2}\;\!\frac{{\rm d}n_{\rm e}}{{\rm d}\gamma_{\rm e}} \int_{0}^{1}\;\!{\rm d}\xi\;\!n_{\rm ph}(\xi) \;\! {f(\xi)}\;\!\xi^{-1} \ ,
\label{eq:ic_emission}
\end{equation}
\citep{Blumenthal1970RvMP} in the Thomson limit. We use the more general Klein-Nishina formula when the Thomson limit is not applicable -- see~\citet{Jones1968PhRv}. Here, $\sigma_{\rm T}$ is the Thomson cross section, $\lambda_c$ is the Compton wavelength of an electron, and we use $f(\xi) = 2 \xi \ln \xi + \xi + 1 - 2 \xi^2$ (for $0<\xi<1$), and the dimensionless variable $\xi = \nu/(4\gamma_{\rm e}^2 \nu')$ where $\nu'$ is the frequency of the target low-energy photon. 
Non-thermal free-free bremsstrahlung CR emission is given by: 
\begin{equation}
    j_{\rm ff}^{\rm nt}(\epsilon) = \lambda_c \nu n_{\rm H} \int_{\sqrt{\epsilon(2+\epsilon)}}^{\infty} {\rm d}\gamma_{\rm e} \;\! n_{\rm e}(\gamma_{\rm e}) \frac{{\rm d}\sigma_{\rm ff}(\epsilon; \gamma_{\rm e})}{{\rm d}\epsilon} \ ,
    \label{eq:ff_emission}
\end{equation}
\citep{Dermer2009_book}, where $\beta \rightarrow 1$ in the relativistic limit, and where we adopt the cross section approximation introduced by~\cite{Baring1999ApJ}. An additional thermal bremsstrahlung component is given by: 
\begin{equation}
    j_{\rm ff}^{\rm t}(\epsilon) = 2 \sqrt{\frac{2}{3\pi \Theta_{\rm e}}}\alpha_{\rm f} c \sigma_{\rm T} n_{\rm H}^2 \exp\left(-\epsilon / \Theta_{\rm e}\right) \epsilon^{-1} G(\epsilon/\Theta_{\rm e})
\end{equation}
\citep{Dermer2009_book},\footnote{Here, $G(\epsilon/\Theta_{\rm e}) = \sqrt{3} \exp(\epsilon/2\Theta_{\rm e}) K_0(\epsilon/2\Theta_{\rm e})/\pi$ is the Gaunt factor in the Born limit, where $K_{0}(\hdots)$ is the modified Bessel function of the second kind, of order zero.} where  
$\Theta_{\rm e}$ retains its earlier definition of dimensionless temperature, but this time is used to denote the temperature of thermal electrons.  $\alpha_{\rm f}$ is the fine structure constant, and we assume a fully ionized thermal gas. We find this component makes only a negligible contribution to the total bubble emission spectrum due to the low gas densities.

\subsection{Post-processing simulations}

We post-process our simulations to compute the resulting emission from an electron spectrum aged to a time that is consistent with the simulated bubble age. {The CR energy densities are computed directly by our MHD simulations and given by $f_{\rm emit} e_{\rm CR}$, which sets the normalization of the CR spectrum (cf. equation~\ref{eq:electron_norm}), where $f_{\rm emit} = 0.003$ (see section~\ref{sec:gal_rad_field}, and \citealt{Yang2017ApJ}). The CR electron spectrum and subsequent emission are computed during post-processing of the simulation output, with 
spectra being modeled at each point throughout the grid~\citep[this is similar to the approach adopted in, e.g.][]{Yang2013MNRAS}. } 
While this post-processing approach cannot strictly capture the true spectral evolution of the CR electrons throughout the simulation domain (for which more sophisticated spectral tracking approaches are required, as that invoked in  e.g.~\citealt{Yang2017ApJ}), we consider it to be a good first-order approximation. This is because the dynamical time of expansion of the bubble is always shorter than the radiative cooling time estimated from our simulations (except for the very early stage of the bubble evolution, which we have accounted for in our spectral ageing treatment in section~\ref{sec:leptonic_bubbles}).

\section{Results}
\label{sec:section4}

\subsection{Bubble spectra}
\label{sec:bubble_spectra}

We calculate the emission spectra from our bubble simulations, ranging from radio frequencies of 50 MHz through to high-energy $\gamma$-rays at 10 TeV. This is intended to cover the expected spectral ranges for the principal emission signatures of the bubbles, 
which would be accessible with current and next-generation instruments, including, e.g. the Square Kilometer Array, SKA in radio bands~\citep{TurnerSKA2014}, \textit{Chandra}\footnote{\url{https://chandra.harvard.edu}} and the up-coming \textit{ATHENA}\footnote{\url{https://sci.esa.int/web/ixo/-/48729-about-athena}} X-ray observatories, and \textit{Fermi}-LAT and the Cherenkov Telescope Array, CTA, in low and high energy $\gamma$-rays, respectively~\citep{Atwood2009ApJ, CTAsciencebook2019}.
The spectra are calculated from the volume emissivity at each point throughout the simulation grid for all processes described in section~\ref{sec:emitted_spectrum} (synchrotron, inverse Compton, thermal and non-thermal bremsstrahlung)\footnote{Our volume emissivity calculations were cross checked against two public codes for synchrotron, inverse Compton and non-thermal bremsstrahlung emission~\citep[this software is described in][]{Hahn2016PoS, Zabalza2016PoS}.}, which are then integrated over the relevant emission volume. We find that, although some thermal bremsstrahlung emission is contributed from regions outside the bubble in our simulations, the contribution from all processes is heavily dominated by that emanating from the bubble itself. Thus we adopt the full simulation domain as our emission volume in computing spectra. We also find that the choice of the initial magnetic field does not have a noticeable impact on our computed spectra. As such, the spectra shown in this section are all computed from our Run A simulation. These are practically indistinguishable from corresponding results from Run B.

\subsubsection{Early evolutionary stage}
\label{sec:early_evo_compare_obs}

\begin{figure}
\includegraphics[width=\columnwidth]{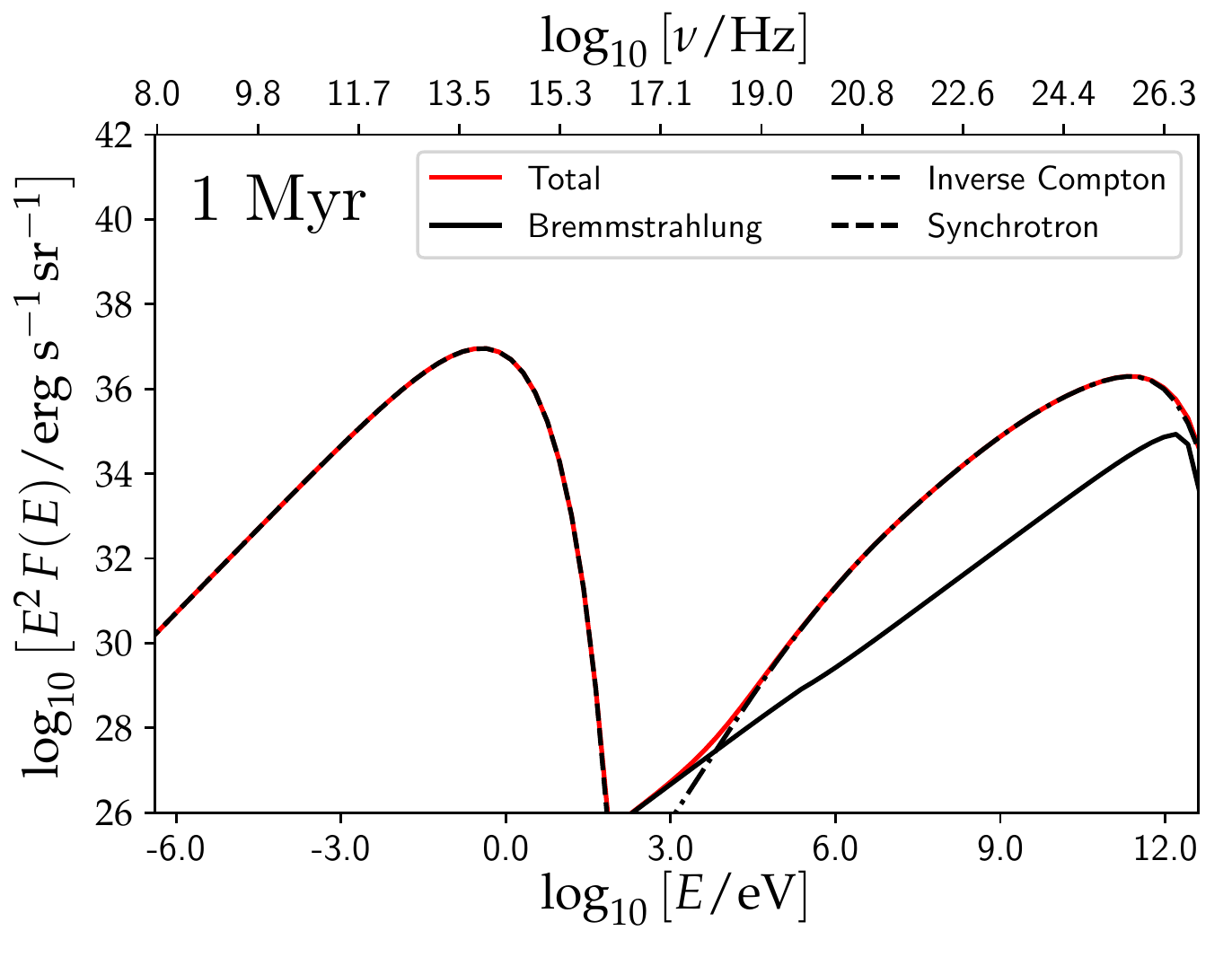}
\caption{Post-processed emission spectrum of the simulated CR bubble (Run A) at 1 Myr. The photon energy is shown in the bottom abscissa, while the corresponding frequency is shown on the top abscissa. The total spectrum (black) can be decomposed into contributing processes, as labeled.}
\label{fig:1myr_total_spec}
\end{figure}

Figure~\ref{fig:1myr_total_spec} shows the bubble spectrum at 1 Myr. The contributions from non-thermal bremsstrahlung, inverse Compton and synchrotron emission are plotted. {We found the thermal bremsstrahlung contribution to the emission was sub-dominant (by a factor of $\sim 10$ at keV energies), and is not shown. However, 
the exact balance between thermal and non-thermal bremsstrahlung emission in the X-ray band is unclear in our post-processed results. The thermal bremsstrahlung emission intensity is particularly sensitive to the simulation resolution, and our calculated values are likely understated. We expect that higher resolution realizations would significantly enhance the level of thermal bremsstrahlung emission from a bubble, and this would likely dominate over non-thermal bremsstrahlung. However, we show in section~\ref{sec:obs_implications} that this is not of great consequence to our present work, where we show that X-ray emission from bubbles around external galaxies would fall far below detection thresholds, and would not practically be detectable whether dominated by thermal or non-thermal bremsstrahlung.} 
It can be seen that the high-energy $\gamma$-ray emission is dominated by inverse Compton emission. This is predominately driven by CR electrons scattering off ISRF photons. At lower energies, up-scattered CMB photons instead make a larger contribution to this emission component. An additional relativistic bremsstrahlung peak can also be seen at TeV energies, but this is sub-dominant during the first $\sim$1 Myr of the bubble's development.

At 1 Myr, the bubble age and structure (see Figure~\ref{fig:bubble_evo}) is  similar to that inferred for the Milky Way's \textit{Fermi} bubbles, and our model spectrum thus show certain similarities to observations. For example,  Figure~\ref{fig:1myr_total_spec} shows the total $\gamma$-ray emission between 1 - 100 GeV is $\sim 10^{36}\;\!{\rm erg}\;\!{\rm s}^{-1}\;\!{\rm sr}^{-1}$, or  $\sim 10^{37}\;\!{\rm erg}\;\!{\rm s}^{-1}$, if assuming isotropic emission, which is consistent with the 1-100 GeV $\gamma$-ray emission 
found for the \textit{Fermi} bubbles
~\citep{Su2010ApJ}. At X-ray energies between 
{$\sim$0.1 - 2 keV, the emission as a total power of $\sim 10^{27}\;\!{\rm erg}\;\!{\rm s}^{-1}$.}
{This is substantially lower than estimated by X-ray observations towards the \textit{Fermi} bubbles by \textit{ROSAT}~\citep{Snowden1997ApJ}, as well as the X-ray luminosity of possibly associated structures (e.g. the \textit{eROSITA} bubbles; see~\citealt{Predehl2020Natur}).
However, these observations include X-ray emission contributed by all the gas in the Milky Way halo, which likely extends to a radius of $\sim 250\;\!{\rm kpc}$~\citep[e.g.][]{Blitz2000ApJ, Grcevich2009ApJ}, the Galactic bulge, and/or features external to the galactic bubbles, which are not present in our model (e.g. including the North Polar Spur;~\citealt{Kataoka2013ApJ}). Additionally, we consider that under-resolved shocks in our simulation may lead to an understatement of their thermal bremsstrahlung contribution.} We find that the radio synchrotron emission in our model has a total power spectral density of $\sim 10^{15}\;\!{\rm W}\;\!{\rm Hz}^{-1}$ at $408$ MHz, 
which is {also} significantly lower than values previously suggested for the \textit{Fermi} bubbles (see, e.g.~\citealt{Jones2012ApJl} which estimated the 408 MHz emission from a similar region using the all-sky data of~\citealt{Haslam1982AAS}). This may result from our simulations insufficiently resolving magnetic field structures around the bubble {surface or draping layer} (see also section~\ref{sec:mhd_comparison_lit}), 
or may indicate the presence of a stronger initial halo magnetic field than that adopted in our simulations. We note that the synchrotron peak emerges at optical wavelengths. However, the total optical synchrotron power, of around $10^{37}\;\!{\rm erg}\;\!{\rm s}^{-1}$, would be approximately 7-8 orders of magnitude less luminous than {starlight of} a typical Milky Way-like host galaxy, and would be diffuse and practically difficult to disentangle from emission associated with a distant host.

\subsubsection{Late evolutionary stage}
\label{sec:late_evo_stage}

\begin{figure}
\includegraphics[width=\columnwidth]{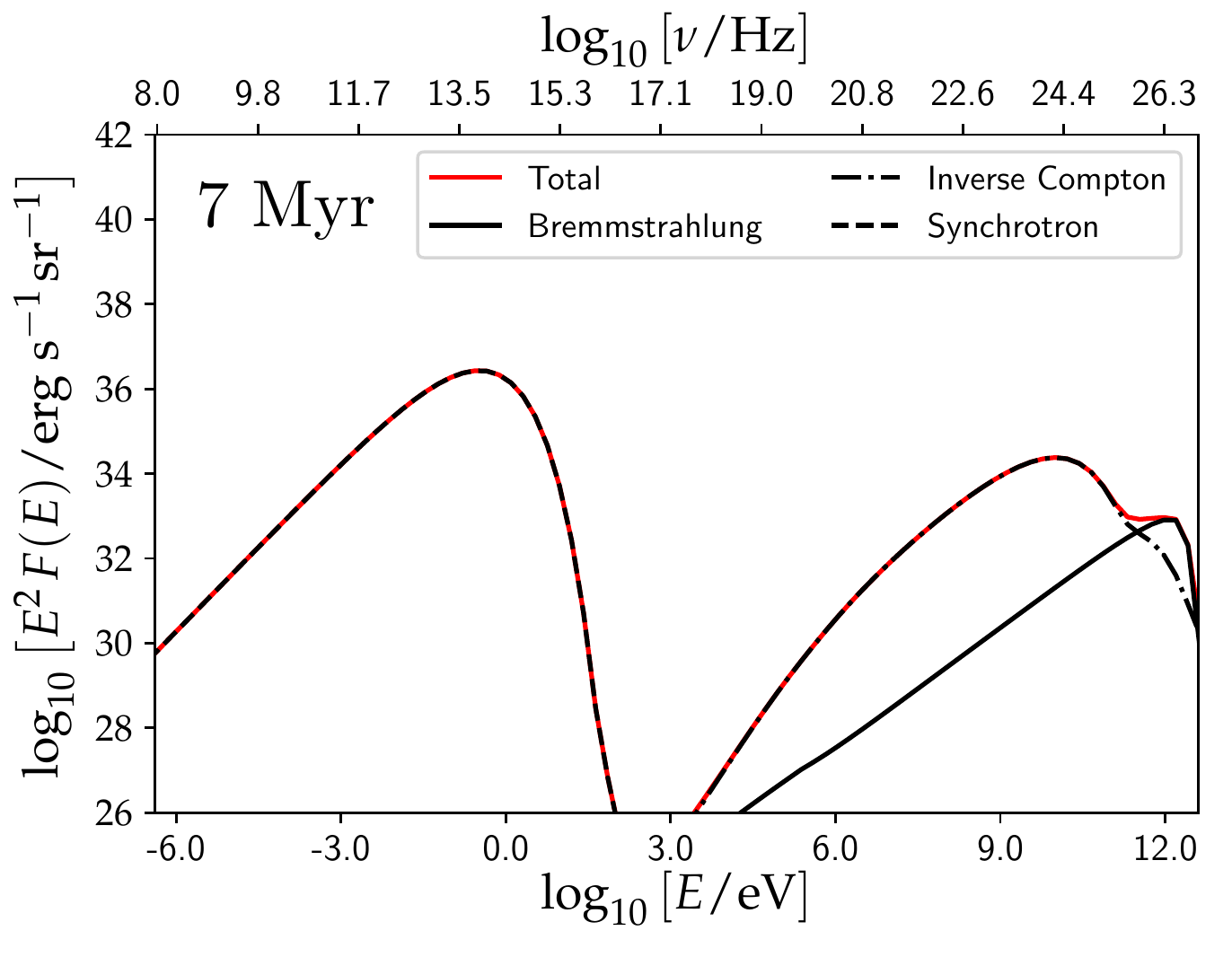}
\caption{Same as Figure~\ref{fig:1myr_total_spec}, but for the post-processed bubble emission spectrum at 7 Myr. }
\label{fig:7myr_total_spec}
\end{figure}

The bubble spectrum at 7 Myr is shown in Figure~\ref{fig:7myr_total_spec}. While similarities to the 1 Myr spectrum remain, there are clear changes in the emitted spectrum as the bubble ages. Overall, emission from all processes reduces between 1 and 7 Myr. This is particularly noticeable at high-energies, where the ISRF inverse Compton component is much weaker at 7 Myr than at 1 Myr (Figure~\ref{fig:1myr_total_spec}), with the spectrum above a few 10s GeV instead being dominated by non-thermal bremsstrahlung. The reduction in the synchrotron emission by 7 Myr is much less severe. In Figure~\ref{fig:bubble_evo}, it can be seen that the CR energy density is much more diffuse at 7 Myr compared to 1 Myr, and the surface magnetic field is comparatively weaker (for discussion, see section~\ref{sec:gen_char}). Although this leads to a substantial drop in synchrotron volume emissivity from the surfaces of the bubble, Figure~\ref{fig:bubble_evo} shows the total emission volume of the surface region is also much greater at later times, largely compensating for this reduction.

\subsubsection{Spectral evolution}

\begin{figure*}
\includegraphics[width=0.7\textwidth]{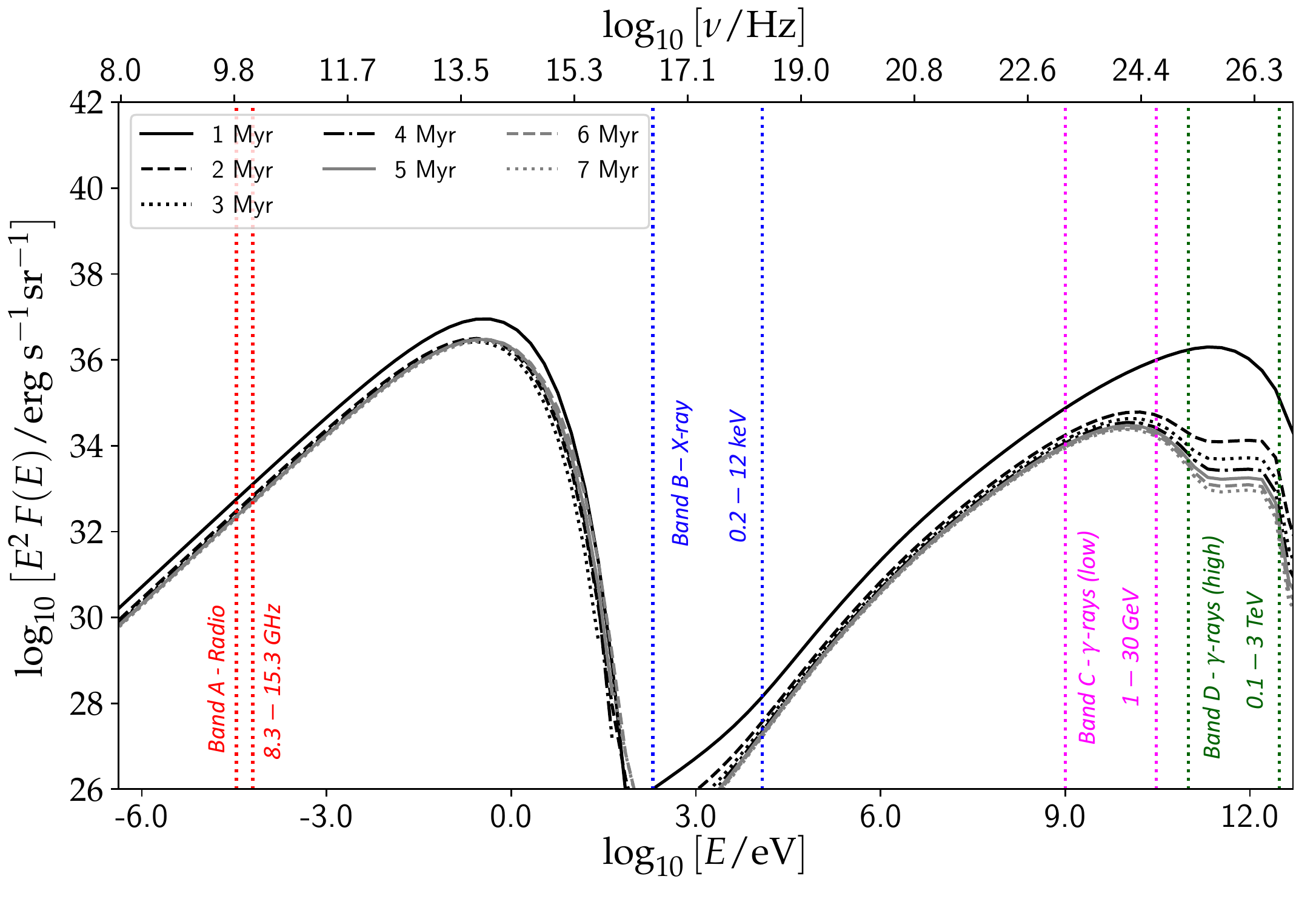}
\caption{Post-processed total emission spectrum of the simulated bubbles between 1 Myr and 7 Myr, with spectra shown in 1 Myr increments as labeled. The photon energy is shown in the bottom abscissa, while the corresponding frequency is shown by the top abscissa. Four spectral bands are marked at radio frequencies (8.3-15.3 GHz, Band A) and at X-ray (0.2-12 keV, Band B) and at $\gamma$-ray energies (1-30 GeV, and 0.1-3 TeV; Bands C and D, respectively), which are used to compute broad-band emission maps in section~\ref{sec:broadband_maps}.}
\label{fig:bubble_spec_evo_all}
\end{figure*}

CR-driven processes dominate the emission from galaxy bubbles. Thus, as CR distributions and the conditions they encounter within a bubble change as it evolves, their emitted spectral properties reflect this. 
Figure~\ref{fig:bubble_evo} shows that the greatest changes in the distribution of CRs throughout our simulated bubbles happens during the first 2 Myr, when the bubble expansion is most rapid. This is reflected in the spectral evolution, shown in Figure~\ref{fig:bubble_spec_evo_all}, where the clearest changes arise over a similar timescale. 
In particular, the large reduction in high-energy $\gamma$-rays noted when comparing Figure~\ref{fig:7myr_total_spec} with Figure~\ref{fig:1myr_total_spec} can be attributed to the evolution of the system in just the first 2 Myr. 
In Figure~\ref{fig:bubble_evo}, it can be seen that the CR energy density becomes more concentrated at the leading edge of the bubble as it ages and expands. At 1 Myr, the bulk of the CR energy density in the bubble is located within a few kpc of the host galaxy, and is subject to a relatively strong ISRF. This yields substantial inverse Compton emission at GeV energies and above. As the bubble expands, much of this CR energy density moves to higher altitudes, where the ISRF is much weaker and, by 3 Myr, the bulk of the energy density and leading edge of the bubble is located well beyond 10 kpc, far beyond the influence of the ISRF. The corresponding inverse Compton emission thus falls quickly during the early evolution of the bubble, but begins to settle later on when it is instead regulated by up-scattered photons from the spatially-uniform CMB.

At later times, the high-energy $\gamma$-ray emission is attributed to non-thermal bremsstrahlung. This process is also responsible for much of the X-ray emission between 0.2 and 10 keV, and ages more gradually than the inverse Compton emission (cf. the change between the 1 Myr and 2 Myr spectra, compared to subsequent reductions in Figure~\ref{fig:bubble_spec_evo_all}). This is because regions of the bubble contributing the most bremsstrahlung emission are where CR energy density and gas density coincide. In our treatment of CRs, where CR energy density and gas are advected together, the CR distribution only varies compared to the gas through the effects of diffusion. While diffusion acts to reduce CR energy density with respect to the gas (thus reducing non-thermal bremsstrahlung emissivity), the timescale of this reduction (the diffusion timescale, $\ell^2/4 \kappa_{||} \sim 20\;\! {\rm Myr}$ over distances of $\ell \sim$10 kpc) is typically much longer than advection timescales governing the global distribution of gas and entrained CRs (of order a few kyr if estimated from the gas velocities indicated in Figure~\ref{fig:velocities}), so changes to the bremsstrahlung emission would be relatively slow.

At low energies, the emission is entirely dominated by synchrotron. This shows very little evolution throughout the 7 Myr bubble lifetime, dropping by just a factor of a few (with a reduction that is effectively energy-independent). Synchrotron emission arises from regions of the bubble where CR energy density and strong magnetic fields coincide. This is almost exclusively attributed to the top of the bubble and, to a lesser degree, the region between the inner and outer contact discontinuities, from which only minimal evolution to the overall emission from the bubble would be expected (for details, see section~\ref{sec:late_evo_stage}).

\subsection{Broadband emission maps}
\label{sec:broadband_maps}

\begin{figure*}
\includegraphics[width=\textwidth]{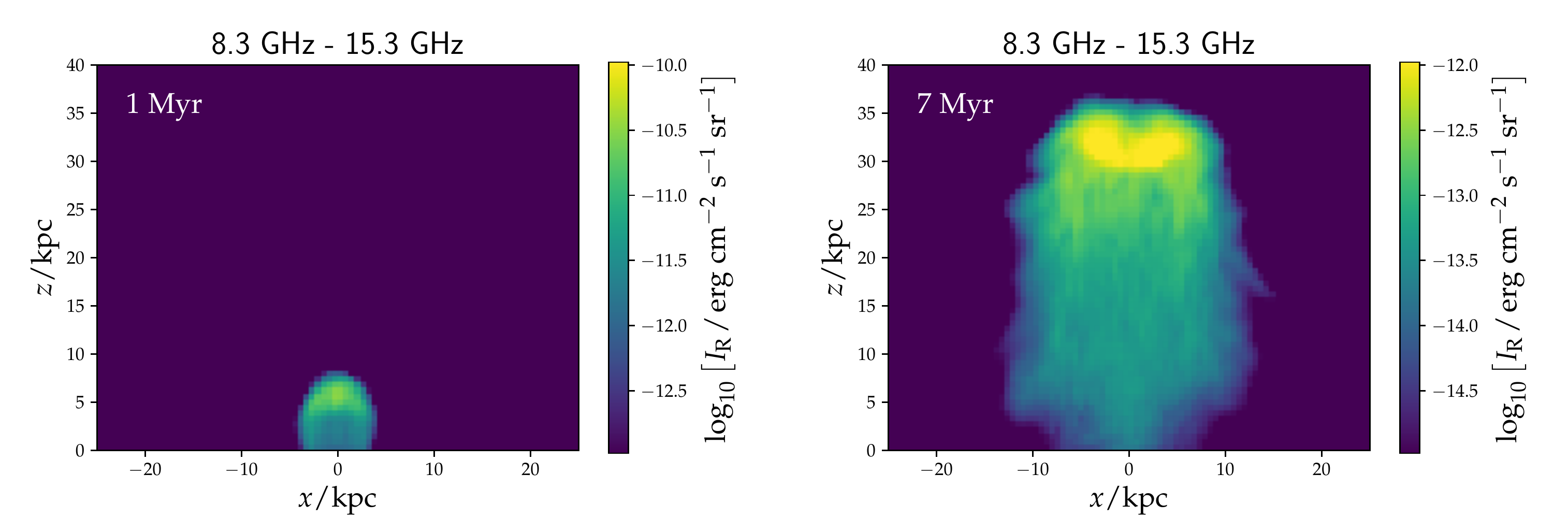}
\caption{2D projection plot of the radio emission from our simulated (Run A) bubble, integrated over 8.3-15.3 GHz (Band A), computed with a post-processing approach. The left panel shows the emission at 1 Myr, while the right panel shows the emission at 7 Myr. The integrated emission over each map is equivalent to the band-integrated emission through Band A shown in Figure~\ref{fig:bubble_spec_evo_all} for corresponding bubble ages. See Figure~\ref{fig:run_b_synch} for the equivalent 7 Myr result for simulation Run B.}
\label{fig:radio_spatial}
\end{figure*}

\begin{figure}
\includegraphics[width=\columnwidth]{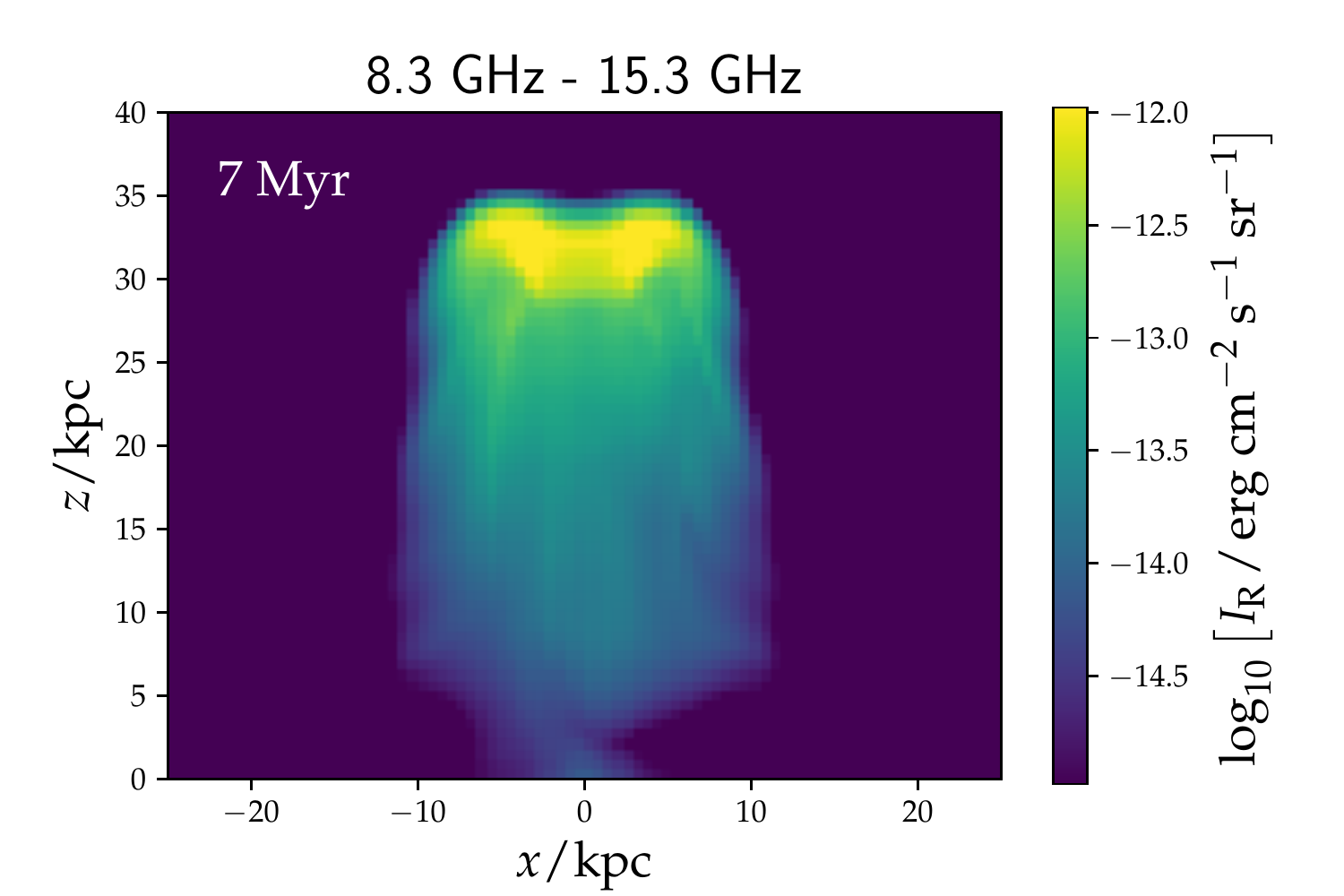}
\caption{{2D projection plot of the radio emission from our simulated Run B bubble at 7 Myr, integrated over 8.3-15.3 GHz (Band A). Comparison with the 7 Myr panel in Figure~\ref{fig:radio_spatial} shows the absence of filamentary features on the bubble surface in the Run B result.}}
\label{fig:run_b_synch}
\end{figure}

Although only observationally accessible in a small number of cases in nearby galaxies and/or for certain wavebands (e.g. radio frequencies), 
broadband spatial emission maps can give useful insight into the origin of spectral signatures, and are informative to connect the emitted radiation calculated for our simulations to their MHD properties. In this section, we show the spatial emission from our bubble simulations in four selected energy bands as marked on Figure~\ref{fig:bubble_spec_evo_all} as Band A (8.3-15.3 GHz), B (0.2-12 keV), C (1-30 GeV), and D (0.1-3 TeV). For all bands, we calculate volume emissivities throughout the simulation grid for all processes described in section~\ref{sec:emitted_spectrum} using the same post-processing technique that was adopted for the spectra in section~\ref{sec:bubble_spectra}. For emission maps, however, instead of integrating over the emission volume to compute a spectrum, we now integrate over the spectrum within the required band (with a spectral resolution of 10 bins in energy),\footnote{We found that use of higher energy resolutions would not noticeably change our results.} and along the $y$-axis of the simulation domain. This yields emission maps as 2D band-integrated projection plots. Note that these 2D projections differ from the 2D slice plots used to present the MHD results in section~\ref{sec:bubble_development} (sliced through the $y=0$ plane), and is more appropriate to reflect the geometry of the projected emission that would practically be observed from bubbles in other galaxies. 

We note that the emission maps shown in this section are for our simulation Run A {(with the exception of Figure~\ref{fig:run_b_synch})}. We found that the emission maps derived from the Run B simulations are generally similar, and only show spatial differences in-line with the variations seen in the MHD results presented in section~\ref{sec:bubble_development}. We explicitly discuss cases where such differences arise.

\subsubsection{Radio (Band A)}

\begin{figure*}
\includegraphics[width=\textwidth]{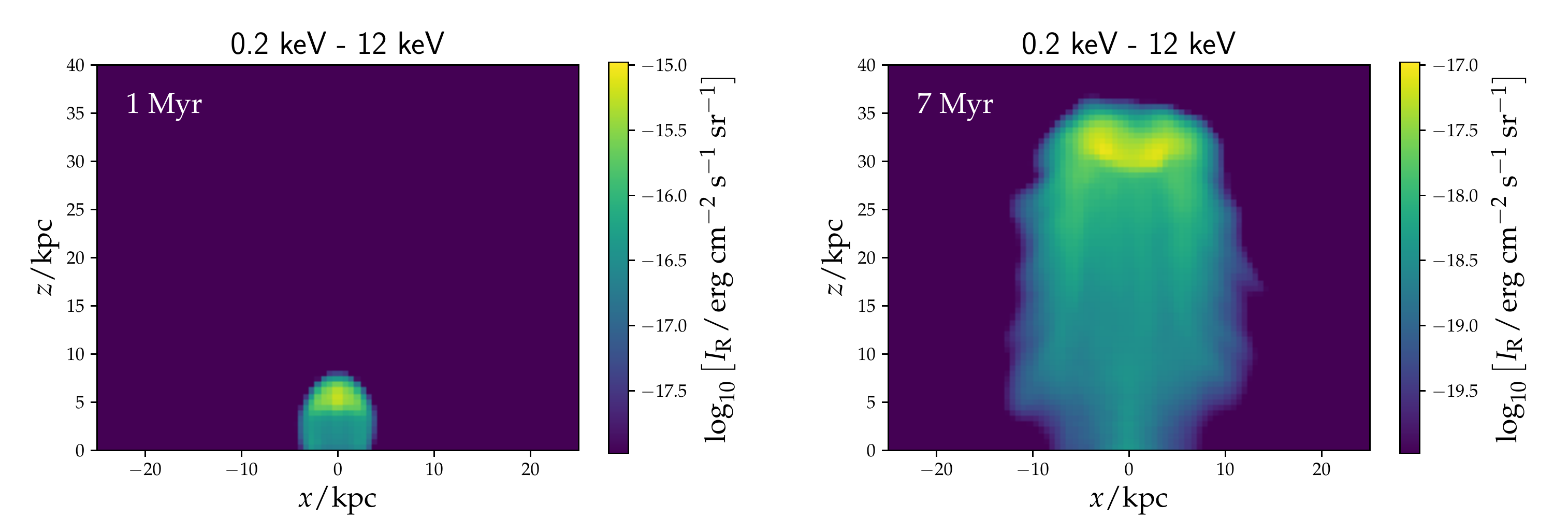}
\caption{2D projection plot of the X-ray emission from our simulated bubble, same as Figure~\ref{fig:radio_spatial}, but integrated over the X-ray energy band B, 0.2-12 keV.}
\label{fig:xray_spatial}
\end{figure*}

\begin{figure*}
\includegraphics[width=\textwidth]{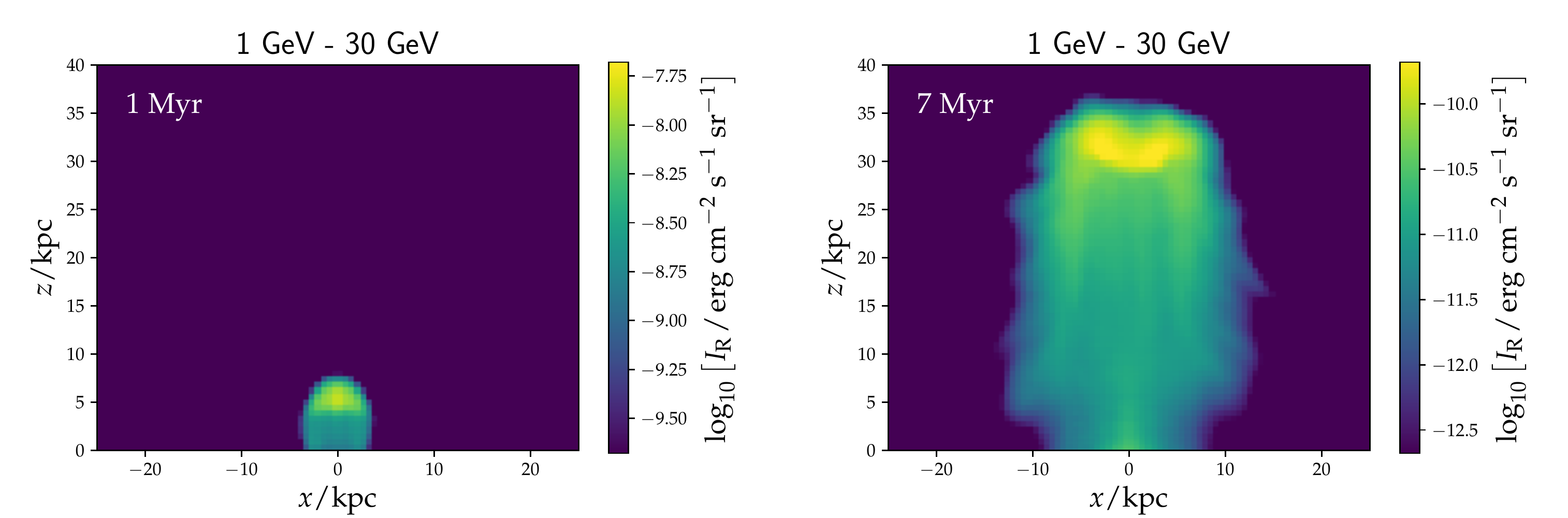}
\caption{2D projection plot of the $\gamma$-ray emission from our simulated bubble, same as Figure~\ref{fig:radio_spatial}, but integrated over the energy band C, 1-30 GeV.}
\label{fig:softgamma_spatial}
\end{figure*}

Figure~\ref{fig:radio_spatial} shows the bubble emission map at radio frequencies, between 8.3 and 15.3 GHz (Band A). At these frequencies, the emission is 
synchrotron-dominated,  
being driven by CR electrons cooling in bubble magnetic fields. At 1 Myr,  the emission preferentially traces the edge of the bubbles (as indicated by an apparent limb brightening in the projected intensity at low altitudes), and 
is also particularly strong towards the top of the structure. This is reflective of the underlying distribution of CRs within the bubble, with emission predominantly arising from regions where a large CR energy density intersects strengthened magnetic fields (e.g. in the draping layer at the bubble edges; cf. Figure~\ref{fig:bubble_evo}). 

By 7 Myr, noticeable differences can be seen to have developed in the structure of the bubble's radio emission:
although the excess at high altitudes remains relatively strong due to the persistent high CR energy density at the bubble leading edge (cf. Figure~\ref{fig:bubble_evo}), much more sub-structure is discernible throughout the bubble. In particular, the edge of the emission region shows a feathered morphology and, across the emission surface, projected filamentary structures can be seen aligned vertically in the prevailing flow direction of the internal gas (cf. Figure~\ref{fig:velocities}).  
Both of these morphological features are a consequence of the magnetic field configuration of the 7 Myr bubble. We note that the edge of the radio emission  corresponds roughly to the position of the outer contact discontinuity, not the edge of the bubble (i.e. at 7 Myr, it is not associated with the draping layer or its remnants; see, in particular, the distribution of the CR energy density in Figure~\ref{fig:bubble_evo}). In this region, the magnetic field is relatively disordered, which leads to uneven CR diffusion near the contact discontinuity. This creates the feather-like emission surface seen in Figure~\ref{fig:radio_spatial}, with brighter filaments emerging within these features, following the structure of the magnetic field. The vertical filaments within the emission region are a consequence of the same effect. These are the face-on feathered emission filaments associated mainly on the bubble's outer contact discontinuity.  
We find these features do not clearly arise in Run B, even at late times {(see Figure~\ref{fig:run_b_synch}, which shows the Run B 7 Myr Band A emission map)}. The persistence of the magnetic draping effect to later times in Run B (see Figure~\ref{fig:magnetic}) leads to a less disordered magnetic field behind the draping layer, including around the contact discontinuity. This forms a smoother emission surface. The implications for the polarized emission properties of the bubble as it ages, and as a consequence of the ambient magnetic field topology, will be investigated in future work (see also, e.g.~\citealt{Yang2013MNRAS}).

We note that the emission intensity is around $10^2$ times lower at 7 Myr than at 1 Myr, a consequence of the bubble's expansion and corresponding reduction in CR energy density. Moreover, at the base of the bubble at 7 Myr, below $\sim$ 3 kpc, the width of the emission appears to be much smaller than the lateral extent of the CR energy density in Figure~\ref{fig:bubble_evo}. While this is primarily a result of our choice of color scale (adopted to highlight the structures at the bubble surface) and low intensity emission does persist to around $x\sim \pm 10$ kpc at the base of the bubble, the emission is  actually somewhat suppressed in this region. This is due to its relatively low magnetic field strength at late times, with magnetic field energy density having been advected away from the lower regions by the rapid and persistent outflowing gas.

Near future facilities offer promising capability to resolve radio bubbles around other galaxies. The frequency range used in our emission maps is representative of the highest energy imaging band planned for the up-coming SKA (in particular, the SKA1-mid array). Imaging observations in this band are expected to be able to achieve excellent angular resolutions ($\sim$0.04'')\footnote{\url{https://www.skatelescope.org}.}, which would allow sub-kpc bubble structures around galaxies to be resolvable out to cosmological distances~\citep{Wright2006PASP}. 

\subsubsection{X-rays (Band B)}

\begin{figure*}
\includegraphics[width=\textwidth]{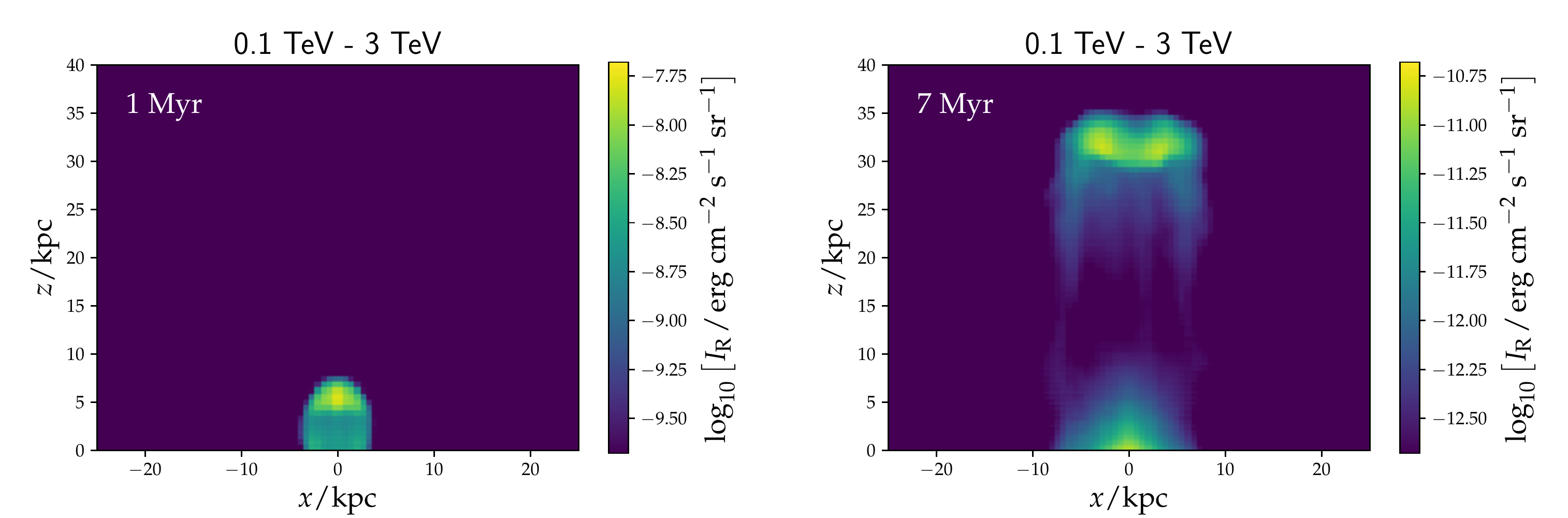}
\caption{2D projection plot of the $\gamma$-ray emission from our simulated bubble, same as Figure~\ref{fig:radio_spatial}, but integrated over the energy band D, 0.1-3 TeV.}
\label{fig:hardgamma_spatial}
\end{figure*}

Figure~\ref{fig:xray_spatial} shows the X-ray emission map, between 0.2 and 12 keV (Band B). This bandwidth is selected to match the energy range over which bremsstrahlung emission dominates the emitted bubble spectrum at all ages (cf. Figure~\ref{fig:bubble_spec_evo_all}), and is 
comparable to the energy ranges over which
current and up-coming X-ray observatories would perform imaging observations, e.g. \textit{ATHENA}, 
which is expected to have imaging capabilities in the 0.5-12 keV energy range with resolutions as high as $\sim$ 5''~\citep[see][]{Willingale2013arXiv}, corresponding to sub-kpc resolution to distances of a few 10s of Mpc~\citep{Wright2006PASP}.

Non-thermal bremsstrahlung emission is strongest in regions where CR energy density and gas density are both high. As such, the low density central lobe (discussed in section~\ref{sec:bubble_development}) is clearly evident in the bubble emission map at both 1 Myr and 7 Myr, manifesting as a lower-intensity emission cavity within the bubble. As with the synchrotron emission, the high CR energy density at the top of the bubble also drives relatively strong bremsstrahlung, while emission intensity also reduces by a factor of $\sim 10^2$ between 1 Myr and 7 Myr (in-line with the reduction in CR energy density between the two epochs). We note that the X-ray emission does not show the same filamentary substructures of the radio emission (comparing Figure~\ref{fig:xray_spatial} with Figure~\ref{fig:radio_spatial}). This is because the hot gas distribution within the bubble is considerably smoother than the magnetic field near the emission surface at the outer contact discontinuity (cf. Figure~\ref{fig:bubble_evo}).

\subsubsection{$\gamma$-rays (Band C and D)}

In Figures~\ref{fig:softgamma_spatial} and~\ref{fig:hardgamma_spatial} we show $\gamma$-ray emission maps for 1-30 GeV (Band C) and 0.1-3 TeV (Band D), respectively. 
These energy bands are chosen to reflect the different underlying physics driving the $\gamma$-ray emission, however they also within reach of current and up-coming facilities, e.g. \textit{Fermi}-LAT at energies between $\sim$ 20 MeV and $\sim$ 300 GeV \footnote{\url{https://fermi.gsfc.nasa.gov}} and the up-coming CTA\footnote{\url{https://www.cta-observatory.org/science/ctao-performance/}} between 0.1 and 10 TeV. 
The resolution of $\gamma$-ray instruments is typically lower than can be achieved in X-rays or at radio wavelengths. However, detections of high-energy bubbles around very nearby galaxies may still be possible~\citep[see also][for possible bubble-like emission structures detected around M31]{Pshirkov2016MNRAS}, and can be complemented with the suite of multi-wavelength capabilities expected to come online in the next decade. 

At GeV energies, $\gamma$-ray emission from the bubble is predominately driven by inverse Compton scattering of CR electrons with CMB photons~\citep[in particular, below 10 GeV -- see also][which find a similar transition between CMB and ISRF inverse Compton seed photons]{Mertsch2011PhRvL}. 
{The CMB photon field} is spatially uniform, so the emission maps in Band C (Figure~\ref{fig:softgamma_spatial}) can be considered a faithful representation of the underlying CR energy density throughout the bubble at all times during its evolution. At higher energies, in Band D, inverse Compton emission is still important, however this is predominately composed of up-scattered ISRF photons rather than those from the CMB. An additional non-thermal bremsstrahlung component can also be important (as indicated earlier, in Figure~\ref{fig:bubble_spec_evo_all}), particularly at late times and high altitudes. In principle, this should lead to emission that is less spatially uniform. However, in Figure~\ref{fig:hardgamma_spatial}, the 0.1 - 3 TeV emission at 1 Myr is actually \textit{more} spatially uniform than in any of the other bands considered. This is because the ISRF inverse Compton contribution is stronger at low altitudes, near the host galaxy, where the interstellar radiation intensity is stronger, while the non-thermal bremsstrahlung emission is more intense near the top of the bubble (where the most CR energy density is located) with some overlap of the two components. By 7 Myr, these have clearly become much more spatially separated and can be easily distinguished: the ISRF inverse Compton emission is concentrated within a $\sim$ 7 kpc radius of the center of the simulation domain while the non-thermal bremsstrahlung emission is clearly visible as a separate component above 30 kpc. 

\subsubsection{{Comment on the modeled early-stage spatial emission structure and observations of the Galactic Bubbles}}

{Figures~\ref{fig:radio_spatial} to~\ref{fig:hardgamma_spatial} show that the computed emission from the top of the simulated bubble is brighter in all energy bands. While this is particularly true in the 7 Myr results, it can also be observed at 1 Myr. The emission structure at 1 Myr should be comparable the Galactic \textit{Fermi} Bubbles, given their similar age. However, observations of the \textit{Fermi} Bubbles show a flat surface brightness in the $\gamma$-ray band, with fainter emission at the top of the bubble in the X-ray band~\citep{Snowden1997ApJ, BlandHawthorn2003ApJ, Zhang2021ApJ}. 
The main reason for this difference is the line of sight projection effect arising from the angle at which we observe the \textit{Fermi} bubbles on Earth. This has non-trivial and significant effects on the emission distribution we observe in the Galactic bubbles. To produce a flat surface brightness, 
emission must increase toward the bubble boundaries/top before projection (see also~\citealt{Yang2012ApJ}). In the X-ray band, bremsstrahlung emission from shocks may also be unresolved (and thus understated) in our simulations, while  
observations include X-ray emission contributed by all the gas in the Milky Way halo, including features external to the \textit{Fermi} bubbles (e.g. the \textit{eROSITA} bubbles; see~\citealt{Predehl2020Natur}), which may lie along the line of sight. These factors are not included in our model (see also the discussion in section~\ref{sec:early_evo_compare_obs}), and could dominate over any remaining surface intensity variations (after projection) in the observations. 
}

\section{Discussion}
\label{sec:section5}

\subsection{Observational implications}
\label{sec:obs_implications}

\begin{table*}
\centering
\begin{tabular}{*{7}{|lc|cccc}}
\hline
{\bf Galaxy} & {\bf Distance/Mpc} & {\bf $t_{\rm A}$ / Myr}  & {\bf $t_{\rm B}$ / Myr}  & {\bf $t_{\rm C}$ / Myr}  & {\bf $t_{\rm D}$ / Myr} 
\\
\hline
M31 & 0.77 & $>$7$^{a}$ & - & 1.5 & 1.4  \\
NGC 4710 & 2.14 & $>$7$^{a}$ & - & 0.8 & 1.0  \\
NGC 3115 & 9.77 & $>$7$^{a}$ & - & - & 0.5  \\
NGC 891 & 10.0 & $>$7$^{a}$ & - & - & 0.4  \\
NGC 4565 & 11.1 & $>$7$^{a}$ & - & - & 0.4  \\
NGC 7457 & 13.2 & $>$7$^{a}$ & - & - & 0.3$^{b}$  \\
NGC 3877 & 14.1 & $>$7$^{a}$ & - & - & -  \\
NGC 3198 & 14.5 & $>$7$^{a}$ & - & - & -  \\
NGC 1386 & 15.3 & $>$7$^{a}$ & - & - & -  \\
NGC 5866 & 15.3 & $>$7$^{a}$ & - & - & -  \\
NGC 3079 & 16.5 & 4.6 & - & - & -  \\
NGC 4388 & 17.1 & 3.2 & - & - & -  \\
NGC 4526 & 17.2 & 3.2 & - & - & -  \\
NGC 7814 & 18.1 & 2.5 & - & - & -  \\
NGC 4013 & 18.9 & 2.2 & - & - & -  \\
NGC 4217 & 19.5$^{c}$ & 1.9 & - & - & -  \\
\hline 
\end{tabular}
\caption{Expected lifetime over which galaxy bubbles would be observationally accessible in Bands A (8.3-15.3 GHz), B (0.2-12 keV), C (1-30 GeV) and D (0.1-3 TeV), if their properties were similar to the bubbles in our simulations, and if located around selected nearby edge-on Milky Way-sized spiral and E/S0 type galaxies within 20 Mpc~\citep[see][]{Li2013MNRAS}.
 We adopt sensitivities of four current/up-coming instruments appropriate for these bands (namely, SKA, \textit{ATHENA}, \textit{Fermi}-LAT and CTA for Bands A, B, C and D, respectively), assuming a 50-hour observation time where appropriate, and expected SKA beam sizes and sensitivities at a nominal frequency of 12.5 GHz was adopted from~\citealt{Braun2019arXiv}. \\
\textbf{Notes}: \\
 $^{a}$ Cases marked with lifetimes of greater than 7 Myr would be observable beyond our simulated evolution time, thus 7 Myr is provided as a lower limit. \\
 $^{b}$ If we estimate that the detectable bubble lifetime is less than 0.3 Myr, we mark it as not detectable. This is because the bubble would still be subject to its initial energetic outburst during this time, and would still be brightening. \\
 $^{c}$ We find bubbles located around host systems at distances beyond $\sim$ 20 Mpc would not be visible in any band.}
\label{tab:observable_times}
\end{table*} 

With the exception of the \textit{Fermi} bubbles, for which there is now abundant data, little is known about the general characteristics of galaxy bubbles. $\gamma$-ray structures similar to the \textit{Fermi} bubbles may exist around other galaxies~\citep[e.g.][]{Pshirkov2016MNRAS}. However, the limited sensitivity and resolution of GeV and TeV observations would preclude the detection of substantial numbers of these phenomena out to great distances, thus limiting our ability to capture the diversity and demographics of galaxy bubble populations. However, in this work we have shown that such bubbles would also leave observational signatures in other parts of the electromagnetic spectrum, where prospects for detection would be substantially better.

In Table~\ref{tab:observable_times}, we consider the scope to detect bubbles around external galaxies in the four emission bands considered in section~\ref{sec:broadband_maps}. 
We adopt sensitivities of four current/up-coming instruments appropriate for these bands (namely, SKA, \textit{ATHENA}, \textit{Fermi}-LAT and CTA for Bands A, B, C and D, respectively) and calculate the time period over which our simulated bubble would remain detectable to these instruments, if located at the positions of selected nearby edge-on Milky Way-sized spiral and E/S0 type galaxies within 20 Mpc~\citep[see][]{Li2013MNRAS}.\footnote{At greater distances, we found a bubble of the configuration considered in this work would not be detectable in any of the bands.}
An edge-on orientation would allow any galaxy bubbles to be discernible from their host, while the type we select is  reflective of those galaxies found to host possible bubble structures in other wavebands, aligned along the minor axis of the host~\citep[e.g.][]{Gallimore2006AJ}.

We find that radio observations offer the best prospect for the detection of galaxy bubbles. 
Table~\ref{tab:observable_times} shows that synchrotron emission would be accessible for much of a bubble's lifetime with SKA~\citep{Braun2019arXiv}. We estimate that structures of similar properties to our simulated bubbles would be detectable up to ages of at least 7 Myr (reflecting the duration of our simulations), at distances of $\sim$ 15 Mpc. However, their slow radio dimming rate in Band A (see Figure~\ref{fig:bubble_spec_evo_all}) would indicate much older bubbles would also be detectable. 
Synchrotron emission from bubbles around galaxies has already been resolved in previous radio observations. While these are not be guaranteed to be the same phenomena as considered here, they do share certain similar characteristics, including the alignment of the extended radio emission with the minor axis of the host galaxy~\citep{Baum1993ApJ, Elmouttie1998MNRAS, Kharb2006ApJ} and, if present, their co-orientation with AGN jets in many spiral Seyfert galaxies where extended galactic-scale radio structures are common~\citep{Gallimore2006AJ}. If these structures are part of the galaxy bubble `family', they could represent systems at a different evolutionary stage and/or subject to different intensities of energy injection and persistence~\citep{Guo2012ApJ}.

Detection prospects in X-rays are particularly poor, even when considering the improved expected sensitivity of \textit{ATHENA}~\citep{Nandra2013arXiv}. Table~\ref{tab:observable_times} shows that none of the bubbles around any of the host galaxies considered would be detectable in X-rays. This perhaps not surprising, given that the bremsstrahlung X-ray flux from our simulated bubbles is several orders of magnitude lower than bubble emission in any other waveband (see Figure~\ref{fig:bubble_spec_evo_all}). Even in the closest of those galaxies considered in Table~\ref{tab:observable_times}, the X-ray luminosity of a bubble would need to be $10^5$ times brighter than considered here to reach detection thresholds.

$\gamma$-rays in Band C (1-30 GeV) also offer relatively limited promise to identify bubbles with 10 years of \textit{Fermi}-LAT data. Only M31 and NGC 4710 are sufficiently close to be visible during the first 1.5 Myr and 0.8 Myr of their evolution, respectively. 
We note that this prediction is consistent with the possible detection of bubbles around M31. Their smaller size of 6-7.5 kpc and slightly higher $\gamma$-ray intensity $4\times 10^{37} \;\!{\rm erg}\;\!{\rm s}^{-1}$~\citep{Pshirkov2016MNRAS} would point towards a younger age than the Galactic \textit{Fermi} bubbles, thus indicating their age would be significantly lower the maximum detectable lifetime of 1.5 Myr estimated here.

Figure~\ref{fig:bubble_spec_evo_all} shows that bubbles would initially be brighter in higher-energy $\gamma$-rays, particularly in first $\sim$ 1 Myr of their evolution. This improves prospects for detection in Band D, if considering CTA's expected sensitivity. 
$\gamma$-ray emission in Band D falls away faster than in Band C. This means that, while more distant galaxies may be detectable, the prospects for observing closer, but older bubbles actually can be \textit{worse} than with \textit{Fermi}-LAT in Band C. For example, bubbles around M31 would fade away at younger ages in higher energies, even though very young bubbles, of ages less than $\sim$0.5 Myr, would be within observational reach of CTA if located around galaxies up to $\sim$ 13 Mpc away.

\subsection{Further remarks}

Our results are sensitive to our simulation set up and initial conditions. These include the configuration of the initial magnetic field (see section~\ref{sec:mag_init_effect}, also~\citealt{Yang2013MNRAS}), the resolution of our simulation (see section~\ref{sec:mhd_comparison_lit}), the energy injected during the initial outburst and its duration~\citep[e.g.][]{Guo2012ApJ, Zhang2020ApJ}, the effect of multiple outbursts, and the ambient conditions surrounding the bubble. Moreover, bubbles in the more distant Universe may evolve very differently to those closer by (e.g. inverse Compton cooling of CR electrons would be more severe at cosmological distances due to the greater energy density of the CMB; this may reduce their ability to drive a bubble's expansion). We consider our results to be sufficient to explore how the leptonic emission properties in different wavebands are connected, how these evolve during the lifetime of a bubble and how they reflect the internal physical properties of the system. However, to draw broader conclusions, and to properly facilitate reliable population modeling, a broader range of physical models and conditions must be considered to properly capture the population diversity of galaxy bubbles.

\section{Conclusions}
\label{sec:section6}

In this work, we investigated the evolutionary properties and emission signatures of leptonic bubbles around galaxies, driven by an initial energetic outburst of energy and CRs. We used 3D MHD simulations that self-consistently include the effects of magnetic fields and anisotropic CR diffusion, as well as the dynamical interaction between the thermal gas of the bubbles and the CRs to investigate their properties and structure as they evolve. We computed the multi-wavelength spectrum from the CR and MHD properties of the bubbles across a broad range of energies, and also showed their spatial emission in four energy bands (radio, X-ray, and GeV and TeV $\gamma$-rays) that capture the spectral components attributed to the dominating physical processes arising within the bubble. We then assessed the observational prospects of such bubbles in these four bands, if they were located in Milky-Way like galaxies at distances of up to 20 Mpc.

We found that the properties of our simulated bubbles and their emission were broadly consistent with those computed in previous simulation work~\citep{Yang2012ApJ, Yang2017ApJ}, while the emission properties of the bubbles at an age of 1 Myr were reflective of those of the Galactic \textit{Fermi} bubbles~\citep[e.g.][]{Su2010ApJ}. We showed that the radio emission of the bubbles was persistent throughout their lifetimes, weakened very gradually, and was driven by direct synchrotron processes between CR electrons and the magnetic fields of the bubble. The X-ray emission was dominated by bremsstrahlung processes, and was spatially reflective of the intersection of the gas and CR energy distribution within the bubble. However the intensity of this X-ray emission was relatively weak and would likely be undetectable in bubbles of this configuration around other nearby galaxies. The GeV and TeV $\gamma$-ray emission was initially dominated by inverse Compton processes between the CR electrons and photons contributed by the CMB and the ISRF of the host galaxy. However, as the bubble expanded, the upper regions harboring the highest CR energy densities moved away from the ISRF. This had a particular impact on the TeV emission from the bubble, which reduced and became dominated by non-thermal bremsstrahlung emanating from near the top of the bubble. We found that the $\gamma$-ray emission remained sufficiently intense for several nearby galaxies to have detectable bubbles of this configuration during the early stages of their evolution, however the high-energy $\gamma$-ray emission would typically fade quickly as the emission transitions from inverse Compton to non-thermal bremsstrahlung.

Bubbles from galaxies in the relatively nearby Universe are not easily detectable at high-energies. However, their associated radio emission is more accessible to current and next-generation instruments. We have demonstrated that self-consistent multi-wavelength emission modeling can provide a connection between radio observations of bubble structures in more distant galaxies with high energy $\gamma$-ray signatures associated with younger bubbles in the nearby Universe. This opens up the possibility of studying a broader range of bubble configurations, initial conditions, compositions and environments to assess the wider demographics of galaxy bubbles and their observational signatures. Such work will be essential to understand the origin and evolution of our own Galaxy’s \textit{Fermi} bubbles, and how they are related to similar phenomena seen further afield.

\section*{Data Availability}

The data generated in this research will be shared on reasonable request to the authors.

\section*{Acknowledgments}

ERO is supported by the Center for Informatics and Computation in Astronomy
  (CICA) at National Tsing Hua University (NTHU)
  through a grant from the Ministry of Education of the Republic of China (Taiwan). HYKY acknowledges support from the Yushan Scholar Program of the Ministry of Education of Taiwan and Ministry of Science and Technology of Taiwan (MOST 109-2112-M-007-037-MY3). 
This work used high-performance computing facilities at CICA, operated by the NTHU Institute of Astronomy. This equipment was funded by the Taiwan Ministry of Education and the Taiwan Ministry of Science and Technology. 
We are also grateful to the National Center for High-performance 
                       Computing (Taiwan) for computer time and facilities. FLASH was developed in part by the DOE NNSA ASC- and DOE Office of Science ASCR-supported Flash Center for Computational Science at the University of Chicago. 
This research has made use of {the {\tt yt} data visualization package~\citep{Turk2011ApJS}, and} NASA's Astrophysics Data Systems. We thank the anonymous referee for their comments.

\bibliographystyle{mnras} 
\bibliography{references} 


\appendix

\section{Cosmic ray spectral evolution}
\label{sec:cr_spec_evo}

We model an ageing electron spectrum in section~\ref{sec:leptonic_bubbles} 
 according to the non-steady distribution function (kinetic equation)
\begin{equation}
    \label{eq:evo_equation_elecs}
\frac{\partial n_{\rm e}(\gamma_{\rm e}, t)}{\partial t} = \frac{\partial }{\partial \gamma_{\rm e}} \left[n_{\rm e}(\gamma_{\rm e}, t)\;\!\beta(t)\right]
\end{equation}
\citep[e.g.][]{Kardashev1962SvA}, where we assume that all particles are injected at $t = 0$ with a power-law spectrum, as given by equation~\ref{eq:electron_spectrum}, and the time-dependent electron cooling rate $\beta(t)$ is dominated by radiative (synchrotron and inverse Compton) and adiabatic cooling,\footnote{While these processes dominate the electron \textit{losses}, the emitted \textit{spectra} (see section~\ref{sec:emitted_spectrum}) from relativistic free-free processes can be non-negligible.}
\begin{equation}
    \beta(t) = \alpha_1(t) \gamma_{\rm e} + \alpha_2(t) \gamma_{\rm e}^2 \ ,
\end{equation}
where $\alpha_1 (t)$ and $\alpha_2 (t)$ are, respectively, 
adiabatic {and} radiative cooling rates, scaled from the cooling rate at a reference energy. In this work, we approximate the adiabatic and radiative cooling timescales experienced by electrons by using fitting functions to emulate the results as evolved in the \textit{Fermi} bubble simulations of~\citet{Yang2017ApJ}. The time-evolving maximum spectral energy is then used as the reference energy at time $t$ to compute effective adiabatic and radiative cooling rates. For times beyond $t_{\rm Fermi}$, the maximum simulation time in~\citet{Yang2017ApJ}, we simply extend the same fitted function for the relevant quantities from $t<t_{\rm Fermi}$ to an appropriate time.\footnote{The exact method to set an appropriate electron cooling function is subjective, however we consider our approach to be reasonable as the resulting spectrum in Figure~\ref{fig:spec_ageing} reflects the general characteristics and upper limit found in~\citet{Yang2017ApJ}, and is able to produce results at 1.2 Myr that are broadly consistent with observations of the Galactic \textit{Fermi} bubbles.}

The evolution equation~\ref{eq:evo_equation_elecs} can be solved (e.g. by method of Greens' functions), to give:  
\begin{equation}
    \frac{{\rm d}n_{\rm e}(\gamma_{\rm e}, t)}{{\rm d}\gamma_{\rm e}} = \mathcal{N}_{\rm e}^{\kappa(\gamma_{\rm e}, t) }\;\! \left( \frac{\gamma_{\rm e}}{\gamma_0} \right)^{-s_{\rm e}\;\!\kappa(\gamma_{\rm e}, t) }  \mathcal{G}(\gamma_{\rm e}, t)\ ,
\end{equation}
where
\begin{equation}
\mathcal{G}(\gamma_{\rm e}, t) = \exp \left\{ 2 A_1(t) \gamma_{\rm e} \int_{0}^t\;\!\frac{\alpha_2(\xi) \;\!{\rm d}\xi}{\kappa(\gamma_{\rm e},t)+ B_1(\xi)}\right\} \ ,
\end{equation}
\begin{equation}
    \kappa(\gamma_{\rm e}, t) = \gamma_{\rm e}\;\!B_1(t) + A_1(t) \ ,
\end{equation}
\begin{align}
    A_1(t) & = \exp\left\{-\int_{0}^t \alpha_1(\xi)\;\!{\rm d}\xi \right\} \ , \\
    B_1(t) & = -\int_{0}^t A_1(\xi)\;\!\alpha_2(\xi)\;\!{\rm d}\xi \ ,
\end{align}
and other terms retain their earlier definitions.


\bsp	
\label{lastpage}
\end{document}